\documentclass[11pt]{article}
\usepackage[T1]{fontenc}
\usepackage[american]{babel}
\usepackage[lmargin=1.1in, rmargin=1.1in, bmargin=1.1in, tmargin=1.35in]{geometry}
\usepackage{hyperref}
\usepackage{xcolor}
\usepackage{float}
\usepackage{graphicx}
\usepackage{amssymb}
\usepackage{amsmath}
\usepackage{enumerate}
\usepackage{mathtools}
\usepackage{booktabs}
\usepackage{enumerate}
\mathtoolsset{showonlyrefs}
\usepackage{ wasysym }
\usepackage[bf,sf]{caption}
\newcommand{\dd}{\text{d}}
\newcommand{\pderiv}[2]{\frac{\partial #1}{\partial #2}}
\newcommand{\deriv}[2]{\frac{\text{d} #1}{\text{d} #2}}
\newcommand{\ppderiv}[2]{\frac{\partial^2 #1}{\partial {#2}^2}}

\newcommand{\eps}{\varepsilon}

\newcommand{\Ltwo}{\mathbb{L}_2}

\newcommand{\Q}{\mathbb{Q}}

\usepackage{mathtools}
\usepackage{epstopdf}

\usepackage[backend=bibtex, doi=false,isbn=false,url=false,
citestyle=numeric-comp , bibstyle=numeric-comp, firstinits=true,
sorting=nyt, maxcitenames=2, maxbibnames=4, hyperref=true, backref=true, eprint=false,
uniquename=false, uniquelist=false]{biblatex}

\renewbibmacro{in:}{}
\newbibmacro{string+doi}[1]{%
  \iffieldundef{doi}{\iffieldundef{url}{#1}{\href{\thefield{url}}{#1}}}{\href{http://dx.doi.org/\thefield{doi}}{#1}}}   
\DeclareFieldFormat*{title}{\usebibmacro{string+doi}{\mkbibquote{#1}}}

\AtEveryBibitem{
 \clearlist{location}
 \clearfield{month}
 \clearlist{address}
 \clearfield{date}
 \clearfield{pages}
 \clearfield{city}
}

\usepackage[]{caption}
\newtagform{brackets2}[\small \textsf]{\textsf{(}}{\normalsize \textsf{)}}
\usetagform{brackets2}

\usepackage[scaled]{helvet}
\usepackage{inconsolata}

\usepackage{titling}
\posttitle{\par\end{center}}
\usepackage{multicol}

\definecolor{dark-red}{rgb}{0.4,0.15,0.15}
\definecolor{dark-blue}{rgb}{0.15,0.15,0.4}
\definecolor{medium-blue}{rgb}{,0,0.5}
\hypersetup{
    colorlinks, linkcolor={dark-red},
    citecolor={dark-blue}, urlcolor={dark-red}
}
\usepackage{lastpage}
\usepackage{fancyhdr}
\usepackage[sf,bf]{titlesec}
\titlelabel{\thetitle.\quad}

\newcommand{\mass}{\mathcal{M}}

\begin{document}

\title{\textbf{ \sffamily  Bidirectionality From Cargo Thermal Fluctuations in Motor-Mediated Transport}}
\author{\sffamily Christopher E. Miles${}^\dagger$, James P. Keener${}^\dagger$\\ \vspace{.1in}{\sffamily \small ${}^\dagger$ Department of Mathematics, University of Utah}}
\date{\sffamily \today}
\maketitle

\renewcommand{\abstractname}{\textbf{\sffamily Abstract}}

\begin{abstract}
Molecular motor proteins serve as an essential component of intracellular transport by generating forces to haul cargoes along cytoskeletal filaments. Two species of motors that are directed oppositely (e.g. kinesin, dynein) can be attached to the same cargo, which is known to produce bidirectional net motion. Although previous work focuses on the motor number as the driving noise source for switching, we propose an alternative mechanism: cargo diffusion. A mean-field mathematical model of mechanical interactions of two populations of molecular motors with cargo thermal fluctuations (diffusion) is presented to study this phenomenon. The delayed response of a motor to  fluctuations in the cargo velocity is quantified, allowing for the reduction of the full model a single ``characteristic distance'', a proxy for the net force on the cargo. The system is then found to be metastable, with switching exclusively due to cargo diffusion between distinct directional transport states. The time to switch between these states is then investigated using a mean first passage time analysis. The switching time is found to be non-monotonic in the drag of the cargo, providing an experimental test of the theory. 
\end{abstract}

\section{Introduction}

Active transport is a key component of cellular function due to the compartmental nature of cellular machinery. This transport is achieved through the use of molecular motor proteins, which undergo a series of conformational changes to walk along cytoskeletal filaments and generate forces to haul cargoes \cite{Howard2001}. The transport of a single cargo can often involve two families of motors that are directed oppositely, denoted \textit{bidirectional transport} \cite{Hancock2014}. For instance, kinesin, which walks in the positive direction of a microtubule, and dynein, which walks in the negative direction, can be attached to the same cargo. Another possibility is that two populations of the same family of kinesin motor can be attached to a cargo but walk along oppositely oriented microtubule tracks \cite{Osunbayo2015}. This phenomenon is observed for a variety of cargoes: mRNA particles, virus particles, endosomes, and lipid droplets \cite{Kunwar2008,Hendricks2010}. Although both families of motors are exerting forces on the cargo in opposite directions, the direction of cargo transport is able to switch. That is, the cargo spends periods of time with a net positive, negative, and zero velocity (denoted a pause state). This distinct switching suggests a mechanism of \textit{cooperation} between the motor families that has been explored from both experimental and theoretical perspectives. 

The role of external influences in the cooperation mechanism remains unclear. A number of studies have identified regulators of kinesin and dynein \cite{Fu2014}. For instance, LIS1 and NudE have been found to modulate dynein's force production capabilities \cite{McKenney2010}. In \cite{ShojaniaFeizabadi2015}, the authors found that the microtubule itself can regulate kinesin force production. However, the necessity of these external regulators for motor coordination in bidirectional transport remains unestablished. The alternative hypothesis relies on the notion that the coordination is a product of the mechanical interactions of the motors with the cargo, denoted a \textit{tug-of-war} scenario.

The tug-of-war hypothesis has also been investigated from a theoretical and experimental perspective. The authors in \cite{Muller2008} formulate the most notable mathematical model capable of producing bidirectionality. In the model, the motors share the load equally. This assumption is not always invoked in later mathematical models. For instance, \cite{Kunwar2011} performs stochastic simulations of unequally distributed motors. However, these authors compare the results of the stochastic simulation with experiments and conclude that switching statistics do not match as the number of motors varies. In \cite{Soppina2009}, another mathematical model is proposed where the two motor populations are required to be asymmetric. That is, the two opposing motor populations must have different force generating properties to break  symmetry. \cite{Lipowsky2006,Lipowsky2010} also provide noteworthy mathematical models, thinking of motor transport as a ``rubber-band''-like process and find rich dynamics. Although not specifically about tug-of-war, motor population models such as the Huxley crossbridge model \cite{HUXLEY1957,Keener2008} use force-velocity relationships for the motor populations. However, since this analysis is a steady-state analysis, it is difficult to infer dynamics, which we address in our model. In \cite{Bouzat2016}, the authors reexamine the mathematical model of \cite{Kunwar2011} and stress the importance of cargo diffusion for the model to produce the right behavior, specifically pointing out the issue of relating steady-state force-velocity curves to dynamics. An asymptotic analysis of a model bearing many similarities to our proposed model (but still with discrete motor number) can be found in \cite{mckinley2012b}. The authors include cargo diffusion in a stochastic differential equation description and note that motor dynamics slow compared to fast fluctuations in the cargo velocity, which ultimately is an important ingredient of our work.

In this work, we present a new tug-of-war model of bidirectional motor-mediated transport. Our proposed model contains fundamentally different essential components than previous work. Broadly, the proposed model is a mean-field model with unequally distributed load. This {differs from} previous discrete motor, unequal load descriptions and therefore requires a different source of noise to induce switching. {By examining the dynamics of a single motor, we quantify the delayed response to instantaneous changes in the cargo velocity, allowing for the use of a force-velocity relationship to infer dynamics. An approximation is then made about how this behavior expands to an ensemble of motors, which leads to the reduction of the system to two ``characteristic distances'', one for each motor population. We verify this approximation is valid with numerical simulations. In the two-variable system,} we find metastability with two states corresponding to positive and negative net velocities, or bidirectional motion. The noise that drives switching between these two states is due to cargo diffusion (thermal fluctuations), an aspect of this process previously noticed but under-emphasized until recently \cite{Bouzat2016}. 

{Previous work has indeed illustrated the significance of motor number fluctuations \cite{Nadrowski2004}. However, in this present work, we choose to use a mean-field model to emphasize the lack of necessity of discrete motor number for bidirectionality. Our proposed  model still incorporates binding and unbinding dynamics and therefore has the same \textit{mean} behavior as a discrete motor model, but lacks the noise associated with discrete events. The only remaining noise source is then cargo diffusion, which we show to be sufficient for bidirectionality. The difference in magnitudes between the fluctuations due to motor number and cargo diffusion is difficult to quantify due to the fundamental difference in structure. In \cite{Guerin2011a}, the authors find that motor number fluctuations can result in an effective diffusion when the number of motors involved in transport is large. }

A characteristic quantity in validating bidirectional transport models is the reversal or switching time of the system: the time between runs of each direction. In our model, the correlation structure of the effect of noise on each population allows for the reduction to an invariant manifold and consequently, a one dimensional mean first passage time problem in a double-well potential. Classical tools can then be used to numerically solve and analytically approximate the corresponding boundary value problem. The switching time is considered as a function of the cargo drag coefficient, which leads to complex behavior as the wells steepen but diffusion strengthens as the drag decreases. Ultimately, the mean switching time is found to be non-monotonic in the cargo drag coefficient, a feature not expected for switching due to motor dynamics. This non-monotonicity provides an experimental test to validate (or refute) our diffusion-driven switching hypothesis.

\section{Methods}

\subsection{Model Formulation}

\begin{figure}[h]
\centering{\includegraphics[width=4in]{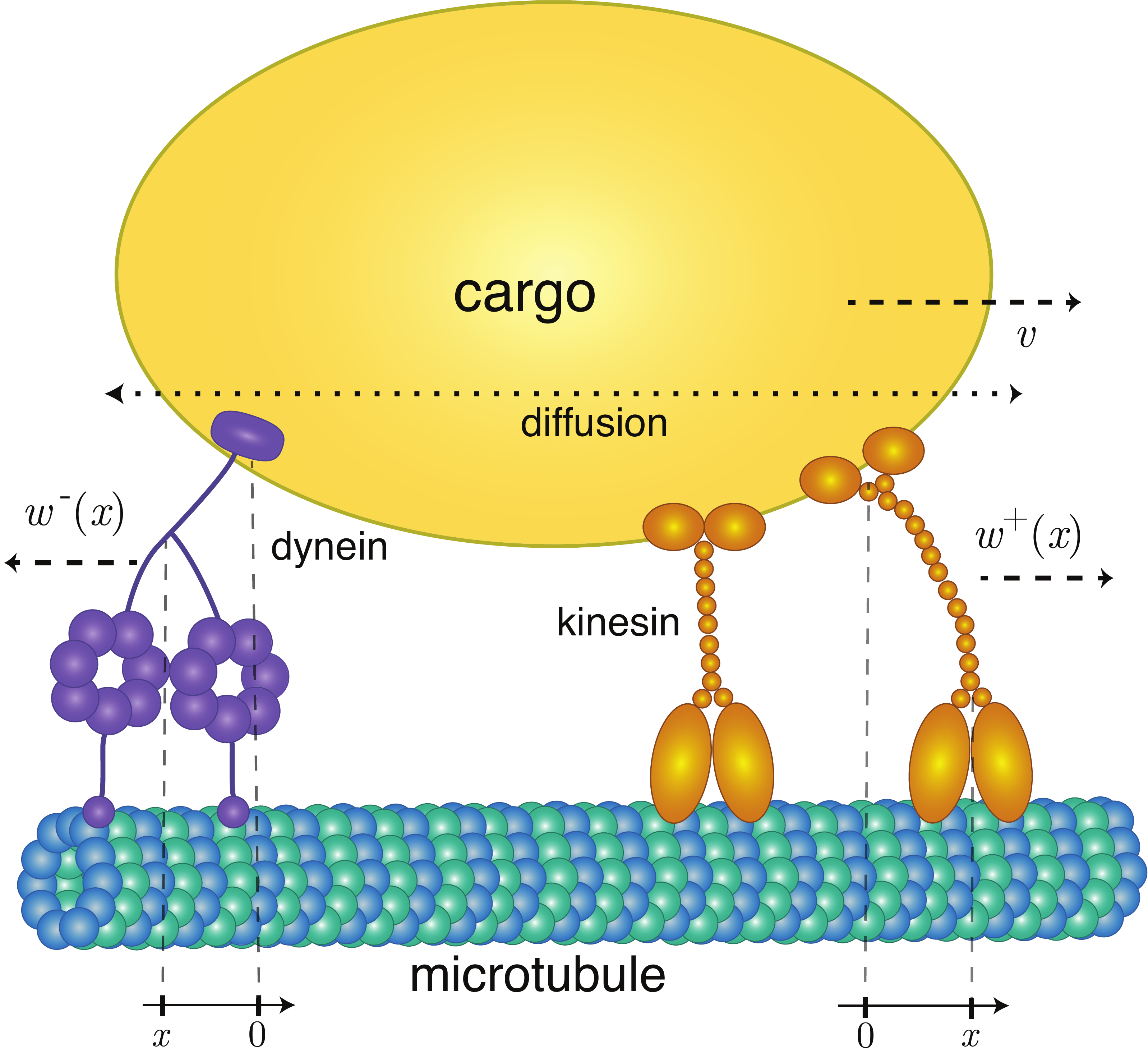}}\\
\caption{A diagram of the mean-field model setup. The quantity, $x$, denoting the distance a motor is stretched is always measured with respect to the orientation of the microtubule.
\label{fig:meanfield}
}
\end{figure}

Consider a cargo being pulled by two different populations of motors, denoted $+$ and $-$. Let $m^\pm(x,t)$ be the density of type $+$ or $-$ motors at time $t$ and stretched from their unstretched distance $x$ units. The $+$ or $-$ labeling of the motor families denotes their preferred directionality. That is, $m^+$ corresponds to the density of motors preferring to walk in the positive direction (e.g. kinesin) and $m^-$ the density of motors preferring to walk in the negative direction (e.g. dynein). The evolution of each motor population is then described by 

{\small 
\begin{align}
\pderiv{m^\pm}{t} + \underbracket{\pderiv{}{x}\left \{ \left[w^\pm(x) - v(t) \right]m ^\pm\right\}}_{\text{stepping}}  =  \underbracket{\left (M^\pm - \int_{-\infty}^{\infty} m^\pm(x,t) \, \dd x \right) {\Omega}^\pm_{\text{on}}(x)}_{\text{binding}} -  \underbracket{{\Omega}^\pm_{\text{off}}(x) m^\pm(x,t)}_{\text{unbinding}}. \label{eq:mean_field}
\end{align}} 

Although \eqref{eq:mean_field} appears as only one equation, $m^+$ and $m^-$ each have their own equation that are structurally identical but may contain different parameters or functional forms. The quantity $x$, describing the distance the motor is stretched from its unstretched displacement is always measured with respect to the microtubule, even though each motor type walks in a different direction, see \textbf{Figure \ref{fig:meanfield}}. This choice of frame of reference is convenient, as it causes the two equations to be structurally identical (as opposed to having to reverse the sign of $v$). 

It is worth noting that this PDE has been studied in other contexts and is referred to as the Lacker-Peskin PDE \cite{Srinivasan2009}, which is an extension of the Huxley crossbridge model \cite{HUXLEY1957,Keener2008}. In that literature, the particular form of the PDE is derived from the limit of a large number of discrete binding sites. {If we consider the case where $M=1$, the mean field model \eqref{eq:mean_field} describes probability density description of a single motor. However, due to the linearity of the equation in $M$, the description of a population of motors is structurally the same but with $M>1$.} 

Before describing, in detail, each term in \eqref{eq:mean_field}, we state a driving assumption for several of the functional forms appearing in the equation. The force generated due to the linker stretching is assumed to be Hookean, that is $\text{force} \sim kx$, where $k$ is the spring constant or stiffness of the motor linker attachment to the cargo. The force-displacement curve of molecular motors has been studied experimentally  \cite{Lindemann2003,Kawaguchi2003} and, although not perfectly linear, seems to be well-approximated by this assumption. 

We now discuss each term of the equation in more detail. Broadly, the motor population can change in three ways: motors stepping (walking), binding or unbinding. 
\begin{enumerate}
\item \textbf{stepping}: We assume that the rate of stepping of the motor is dependent on the force exerted on the motor, which is some function of the distance between the motor and the cargo based on the Hookean force assumption previously mentioned. The walking rate $w(x)$ is therefore linker-distance dependent. We  take the particular functional form
\begin{equation}
w(x) \coloneqq -ax + b,
\end{equation}
where $a>0$. At $x=0$, which corresponds to the motor being unstretched, the motor walks with some velocity $b$. For the $+$ directed motor, for instance, $b>0$. As the motor walks farther from its unstretched position ($x>0$), the force exerted on it causes the velocity to decrease until it eventually stalls at $x_{\text{stall}} \coloneqq b/a$. If $x<0$, that is, the cargo is ahead in the direction the motor seeks to walk, the velocity is assumed to be greater as the linker exerts a force in the direction of motion of the motor. If $x> x_{\text{stall}}$, then the force exerted by the linker is greater than the stall force, meaning the motor moves opposite its preferred direction. 

This force-velocity curve has been qualitatively observed experimentally for kinesin \cite{Kunwar2008,Gennerich2007} and dynein. \cite{belyy2014} and a (non-linear) version of this functional form has been used in a number of modeling papers \cite{Muller2008,Kunwar2011,Bouzat2016}. Although motors (particularly dynein) have non-linear force dependent velocities, we assume that the motors operate within the linear regime. The main deviation from linearity in experimental observed values occurs at superstall forces, where the motor velocities remain negative (as in this model), but with a much smaller magnitude. Since the superstall velocities are larger in magnitude by our approximation, the motors relax back to stall faster and therefore generate a smaller force due to displacement. This is offset by the motors unbindinding less rapidly (due to a lower force), and consequently the net force generated is be approximately the same.

\item \textbf{binding}: The functional form of the binding term is set to be
\begin{equation}
{\Omega}_{\text{on}}(x) \coloneqq k_{\text{on}} \, \, \delta(x),
\end{equation}
where $k_{\text{on}}$ is the constant describing the rate of binding of a molecular motor to the cargo. The $\delta(x)$ functional form corresponds to the assumption that motors are initially unstretched ($x=0$) when they bind, thus only binding at $x=0$. That is, the motors only bind in a non-force-producing state. This assumption can be relaxed (and is for later numerical simulations) to a Gaussian approximation of the delta function.
\item \textbf{unbinding}: The unbinding rate of molecular motors has experimentally been found to be related to the force exerted on them \cite{Kunwar2011,Kawaguchi2003}, however the nature of this dependency is complex and varies from motor to motor. Dynein is found to a have a catch-bond behavior \cite{nicholas2015,Kunwar2011}. Both kinesin \cite{andreasson2015} and dynein \cite{nicholas2015} have been observed to have asymmetric force dependence in their unbinding.

Due to the complexity and variation in unbinding dependence, we take the simplest form that still behaves in a way that qualitatively matches experimental results, which is
\begin{equation}
{\Omega}_{\text{off}}(x) = k_{\text{off}} \, \, \exp \left\{ \frac{k|x|}{F_D} \right \}, 
\end{equation}
where again, the force exerted is assumed to be Hookean $(\sim kx)$, {and independent of direction  (hence the absolute value)}. $F_D$ is a characteristic force fit to experimental observations, and $k_{\text{off}}$ is the unstretched detachment rate. This form is often referred to as Bell's Law, which is known to need corrections in some scenarios \cite{walcott2008load}. The overall behavior of this function establishes that motors detach at a faster rate the farther they are stretched due to the force exerted on their microtubule binding sites.  

This functional form (or similar) has been used in other motor population models \cite{walcott2008load,Srinivasan2009}. In \cite{Kunwar2011}, the authors account for the stalling of motors and the catch-bond behavior of dynein by taking a non-monotonic dependence on the force. In our unbinding rate, neither the catch-bond behavior nor is the asymmetric dependence on force is included. The consequence of excluding these phenomena is purely quantitative, as they are not dramatic enough effects (in the regimes that motors operate for transport) to produce a qualitative effect in our model.

It is also worth noting that $\Omega_{\text{off}}$ and $\Omega_{\text{on}}$ have different units, as the off-rate is multiplied by $m$, a motor density and the on-rate is multiplied by a total number of motors $\int m \, \dd x$. 
\end{enumerate}

We then can define the average force exerted by each motor population, recalling the assumption of a Hookean force,
\begin{equation}
F^{\pm}(t) \coloneqq \int_{-\infty}^\infty k^\pm x m^\pm(x,t)\, \dd x. \label{eq:instantF}
\end{equation} 
This time-varying quantity requires knowledge of the full density of motors $m(x,t)$, which makes it difficult to study directly.

\subsection{Steady-State Analysis}

This time-dependent force, described by \eqref{eq:instantF} is difficult to compute in practice, so we turn our attention to the steady-state force.  We consider the steady state ($\dd m^\pm / \dd t =0$) and behavior of \eqref{eq:mean_field} with some steady-state velocity $\tilde{v}$, which leads to the pair of equations for the steady state densities $\tilde{m}^\pm$
\begin{align}
\pderiv{}{x}\left \{ \left[w^\pm(x) - \tilde{v} \right]\tilde{m} ^\pm\right\}  =  \left (M^\pm - \int_{-\infty}^{\infty} \tilde{m}^\pm(x) \, \dd x \right) {\Omega}^\pm_{\text{on}}(x)-  {\Omega}^\pm_{\text{off}}(x) \tilde{m}^\pm(x). \label{eq:mean_field_ss}
\end{align}

Exploiting the linearity of \eqref{eq:mean_field_ss}, along with the partitioning nature of the delta function, \eqref{eq:mean_field_ss} can be solved analytically, resulting in a solution with an integrable singularity at the stall distance dependent on the velocity
\begin{equation}
x_{\text{stall}} \coloneqq \frac{b-\tilde{v}}{a}.
\end{equation} 
For details of this calculation, see \textit{Supplementary Section S1}. This allows us to define the steady state force exerted by each population of motor
\begin{equation}
\tilde{F}^\pm(\tilde{v}) \coloneqq \int_{-\infty}^{\infty} k^\pm x \tilde{m}(x; \tilde{v}) \, \dd x, \label{eq:Ftilde}
\end{equation}
where we parameterize this force as a function of the steady state cargo velocity $\tilde{v}$ which appears in \eqref{eq:mean_field_ss}.

We now need an equation governing the cargo velocity, which is determined by the forces exerted on the cargo
\begin{equation}
{\mathcal{M}}\dot{v} + \gamma v = \sqrt{2 \gamma k_B T} \xi(t) + \text{forces exerted by motors}. \label{eq:v_orig}
\end{equation}
{In \eqref{eq:v_orig}, $\mathcal{M}$ is the mass of the cargo}, $\gamma$ is the drag coefficient of the cargo and $\xi(t)$ is the white-noise process due to thermal fluctuations (diffusion) of the cargo which satisfies $\langle \xi(t)\xi(\tau)\rangle = \delta(t-\tau)$. The magnitude of these fluctuations is determined by the fluctuation-dissipation theorem \cite{Gardiner2009}. 

\subsection{Forces Exerted By Motors}

{A perhaps natural choice for the force terms in \eqref{eq:v_orig} could be be the steady-state force, $\tilde{F}^\pm(v)$, found in \eqref{eq:Ftilde}. The use of a \textit{force-velocity} relationship (which $\tilde{F}$ is) to study motors has a long history (e.g. \cite{HUXLEY1957}) but there is a problem with this choice. Although $v$ is changing instantaneously, the position of the cargo is not. The forces exerted by the motors are due to stretching of the linker (determined by their displacement), which does \textit{not} change instantaneously as the velocity changes. In other words, it is impossible to completely infer \textit{dynamics} from a \textit{steady-state} force-velocity relationship. Thus, parameterizing the force with time-varying velocity would not produce the physical behavior we desire. For this reason, we turn to a simpler model to understand what to use for the force terms in \eqref{eq:v_orig} that accounts for this issue. }

{In \cite{Bouzat2016}, the authors make the observation that including cargo noise produces this described difficulty: motors should not react instantaneously to velocity and classical models produce results inconsistent with experimental observations if this is the case. To overcome this issue, the authors hypothesize that the motors respond to a time-windowed-average force, suggesting some ``memory'' property of the motors. Here, we directly compute a physiological, mechanistic delay stemming from the stepping of the motor, instead of a phenomenological ``memory''. }

\subsubsection{Ornstein-Uhlenbeck motivation}

{To understand motor response to fluctuations in the cargo velocity, we now examine the behavior of a \textit{single motor} on a \textit{single run}: after binding and before unbinding. Let $x_1(t)$ be the random process describing the distance stretched the single motor is from its unstretched distance and $p_1(x,t)$ be the probability density of this random variable. The behavior follows almost identically with the mean-field model \eqref{eq:mean_field}, but now binding and unbinding can be neglected due to the analysis only being of a single run. In \cite{mckinley2012b}, the authors also study the behavior of motors without binding dynamics and find that multiple motors can actually produce a lower cargo velocity than a single motor. However, we are only interested in the mean behavior of the motors and therefore the explicitness of their description is not necessary for this work. }

{Thus, since this corresponds to a single run of a motor, the only remaining dynamics are the motor stepping (still at its force dependent velocity $w$) and diffusion (the magnitude of which is lumped into a parameter $D$). The resulting process is an Ornstein-Uhlenbeck process \cite{Gardiner2009},  which can be described by the Langevin equation}

\begin{equation}
{\dot{x}_1 = \left [ w(x_1) - v(t) \right] + \sqrt{2D} \, \xi(t)},
\end{equation}
or the corresponding Fokker-Planck equation
\begin{equation}
\pderiv{p_1}{t} = - \pderiv{}{x} \left \{ [w(x) - v(t)] p_1 \right\} + D \ppderiv{p_1}{x}.  
\end{equation}
{To quantify the motor's ability to respond to instantaneous fluctuations in the cargo velocity, we consider the mean value of this single motor, single run process, denoted $\mu_1$,}
\begin{equation}
\mu_1 \coloneqq \langle x_1(t) \rangle. 
\end{equation}
From the Fokker-Planck equation, we find the relationship describing the temporal evolution of the mean of this process to be (assuming $w$ is a linear function)
\begin{equation}
\dot{\mu}_1 = w(\mu_1) - v(t). \label{eq:mueq}
\end{equation}
For details of the calculation, see \textit{\ref{sect:A:OU}}. 

{However, again recalling the assumption of a Hookean force (that is, $\text{force} \sim kx_1$), the average force exerted by a single motor under evolving under this process with density $p(x_1,t)$ is then}
\begin{equation}
 F_{OU}  = k \int_{-\infty}^{\infty} x_1 p(x_1,t) \, \dd x_1 = k \mu_1 \label{eq:FOU}. 
\end{equation}
{In other words, \textit{for a single motor, on a single run}, the force exerted can be parameterized by the mean distance stretched of the motor $\mu$, where $\mu$ ``tracks'' the velocity through \eqref{eq:mueq}, which specifically specifies that the magnitude of the delay is determined by the motor velocity. In other words, changes in force are only due to changes in displacement, not velocity. This resolves the aforementioned issue about the force changing instantaneously. Now, the force tracks, with some delay as determined by \eqref{eq:mueq}, the velocity and evolves continuously.  }

\subsubsection{Force Evolution Approximation}

{The previous calculation showed that while still attached, the force generated by individual motors \eqref{eq:FOU} track instantaneous fluctuations in cargo velocity with a delay related to their processivity, described by \eqref{eq:mueq}. In other other words, the force generation for a population of motors could be collapsed down to a single parameter $\mu_1$. We now make the major approximation of the paper that even with binding and unbinding, the force generated by each population of motors can be collapsed to a single parameter $\mu$ (for each population) with a similarly structured delay. This leads us to the set of equations}

\begin{equation}
{\mathcal{M}} \dot{v} + \gamma v = \hat{F}(\mu)  + \sqrt{2 \gamma k_B T} \xi (t), \qquad \dot{\mu} = w(\mu) - v. \label{eq:mv}
 \end{equation}

{Thus, the cargo velocity $v$ evolves with the forces exerted on it, but the force exerted by the motors is not directly prescribed by the current $v$ but rather some parameter $\mu$ which tracks $v$ with a delay. We can regard $\mu$ as a ``characteristic distance''. That is, the force exerted by each population of motors is entirely parameterized by some dynamic variable $\mu$, determined by $\hat{F}(\mu)$. This could be thought of as effectively the force exerted by a population of motors with mean displacement $\mu$, staying in the spirit of the mean-field model.  It is important to note that \eqref{eq:mv} is written for a single $\mu$, meaning a single motor population to demonstrate the structure but we later incorporate a $\mu_1, \mu_2$, one for each population. We have also not yet specified the choice of $\hat{F}$ but rather are illustrating the structure of the dynamics. }

{Although the motivation for \eqref{eq:mv} was in the motor-attached scenario, the particular choice of $\hat{F}_j(\mu)$  (for  $j=1,2$, corresponding to each population) must not neglect unbinding and binding of the motors incorporated in the mean-field model.} Thus, we take the force exerted by the motors to be the steady state force generated, described by \eqref{eq:Ftilde} such that
\begin{equation}
\hat{F}_{j} (\mu_j) = \tilde{F}_j(-a_j \mu_j + b_j ). \label{eq: hatF}
\end{equation}
We justify this by observing that $\tilde{F}$ was computed for motors equilibrated for a particular constant $\tilde{v}$, which we can think of this as when $\dot{\mu} =0$ (that is, the mean displacement of the population is not changing), and therefore
\begin{equation}
\dot{\mu} = 0 = -a \mu + b - \tilde{v} \implies \tilde{v} = -a \mu + b,
\end{equation}
meaning we associate $-a_j \mu_j + b_j$ with $\tilde{v}$ to obtain \eqref{eq: hatF}. In other words, the motors track the steady state force-velocity curve $\tilde{F}$ with some delay. {This particular choice of the force structure allows for the complexity of the mean-field model, including all binding and unbinding to be embedded into the $\tilde{F}(\mu)$ terms. However, the dynamics of the reduced ``characteristic distance'' model are easier to study due to being an ordinary differential equation rather than a partial differential equation. }

\subsection{Full Model}

The parameter regime we are considering deals with cargo with negligible mass, thus suggesting we are in a viscous or near-viscous regime. Exploiting this fact, we can perform an adiabatic (quasi-steady state) reduction  on \eqref{eq:mv} to eliminate $v$. For details of this calculation, see \textit{Supplementary Section S3}. The result of performing this reduction is
\begin{equation}
\dot{\mu} = w(\mu) - \frac{\hat{F}(\mu)}{\gamma} + \sqrt{\frac{2 k_B T}{\gamma}} \xi(t), \label{eq:mudot}
\end{equation}
or equivalently, in Fokker-Planck form
\begin{equation}
\pderiv{p}{t} = - \pderiv{}{\mu} \left \{ w(\mu) - \frac{1}{\gamma} \hat{F}(\mu) \right\} + \frac{k_B T}{\gamma} \ppderiv{p}{\mu}. \label{eq:mufp}
\end{equation}

One important note from the calculation detailed in \textit{Supplementary Section S3} is that although $v$ is eliminated from the system, $v$ relaxes quickly to a Gaussian centered around 
\begin{equation}
	\hat{v} \sim \hat{F}(\mu)/\gamma, \label{eq:hatv}
\end{equation} 
thus the value of $\mu$ directly determines the (mean) velocity of the cargo at any time.

Combining all of the previous observations, we now propose the full model. In the derivation of \eqref{eq:mudot},\eqref{eq:mufp}, only one motor population was considered, but in bidirectional transport, there are two populations evolving separately, resulting in two equations with identical structure but different parameters.  From this, we get the full model
\begin{align}
\begin{aligned} \label{eq:2ODE}
\dot{\mu}_1 &= -a_1 \mu_1 +b_1  - \frac{1}{\gamma}\left\{F_1(\mu_1) +F_2(\mu_2)\right\} + \sqrt{\frac{2 k_BT }{\gamma}}\xi(t),\\
\dot{\mu}_2 &= -a_2 \mu_2 +b_2  - \frac{1}{\gamma}\left\{F_1(\mu_1) +F_2(\mu_2)\right\} + \sqrt{\frac{2 k_BT }{\gamma}}\xi(t).
\end{aligned}
\end{align}
Note that we have switched the two populations to labels $j=1,2$ instead of $+/-$ for notational convenience. We have also used the functional form of the motor force velocity curve $w(x) = -ax+b$ and that the net force exerted by the motors is simply the sum of the force exerted by each population.

\begin{table}[h] 
{\footnotesize  \centering
\begin{tabular}{cccccccc}
\hline
$F_{stall}$ {\footnotesize $[\text{pN}]$} & $v_0$ {\footnotesize  $\left[\text{nm}\cdot \text{s}^{-1}\right]$} & {$k_\text{off}$} {\footnotesize $\left[\text{s}^{-1}\right]$} & $|F_d|$  \, {\footnotesize $\left[\text{pN}\right]$}  & {$k_{\text{on}}$}   {\footnotesize $\left[ \text{s}^{-1}\right]$} & $M$ & $k$ \, {\footnotesize $\left[\text{pN}\cdot \text{nm}^{-1}\right]$} & $\gamma$  \,  {\footnotesize $\left[\text{pN}\cdot \text{s} \cdot \text{nm}^{-1}\right]$}  \\ \hline
5 & 1000 & 1 & 1 &  5 & 10 & 0.4 & 0.001  \\ \hline
\end{tabular} }
\caption{``Typical'' motor values used for both populations of motors in the symmetric case of the  mean field model. {Values used are within reported ranges of kinesin and dynein.} }
\label{tab:params_general}
\end{table}

To emphasize the ability of this model to produce bidirectional motion without asymmetry  between the motor populations, we take the parameters describing each of the populations to be the same (unless noted otherwise), described in \textbf{ Table \ref{tab:params_general}}. {These parameters are chosen as physiologically reasonable parameters in the range of reported values of both kinesin and dynein, taken from \cite{Kunwar2008,Schnitzer2000,Klumpp2015}. The viscosity of cytoplasm is reported to be higher than water \cite{Mitchell2009,LubyPhelps2000}. Although a potentially large viscosity is used in this work, any smaller would only make the magnitude of the fluctuations larger, further magnifying the importance of cargo diffusion.}

\subsection{Dimensional Reduction}

An important observation must be made about the noise structure of \eqref{eq:2ODE}: the white noise term in each equation is exactly the same (fully correlated). From a biophysical perspective, this is because the two motors feel the same fluctuations from the cargo diffusion. Hence, this is truly a one-dimensional diffusion rather than two dimensional as it currently appears. The Fokker-Planck equation corresponding to the system \eqref{eq:2ODE} has a non-invertible diffusion tensor, which further illustrates this point.  To make the one-dimensional structure more apparent, we perform a change of variables, taking
\begin{equation}
	\zeta \coloneqq \mu_1 + \mu_2, \qquad \eta \coloneqq \mu_1 - \mu_2 \qquad \implies \qquad \mu_1 = \frac{1}{2}(\eta+\zeta), \qquad \mu_2 = \frac{1}{2}(\eta - \zeta).
\end{equation}
Under this coordinate change, the system \eqref{eq:2ODE} becomes, abbreviating $D \coloneqq k_BT/\gamma$
\begin{align}
	\dot{\zeta} &= -\frac{a_1}{2}(\zeta+\eta)+b_1 - \frac{a_2}{2}(\zeta-\eta)+b_2 - \frac{2}{\gamma} \sum F + 2\sqrt{2D} \xi(t) \\
	\dot{\eta} &= -\frac{a_1}{2}(\zeta+ \eta) + b_1 + \frac{a_2}{2}(\zeta-\eta) - b_2. \label{eq:rotated}
\end{align}
By taking the two populations to be symmetric, which corresponds to $a_1 = a_2 = a$ and $b_1 = -b_2 = b$, the $\eta$ equation becomes
\begin{equation}
\dot{\eta} = -a \eta + 2b,
\end{equation}
which has an invariant manifold described by $\tilde{\eta} = 2b/a$. Since the equilibria of the system must lie on this invariant manifold, all dynamics of interest evolve on the manifold and consequently reduces the problem to the one-dimensional evolution 
\begin{equation}
		\dot{\zeta} = -a\zeta - \frac{2}{\gamma}\left[F_1\left(\frac{\zeta+\tilde{\eta}}{2}\right)+F_2\left(\frac{\zeta-\tilde{\eta}}{2}\right) \right] + 2\sqrt{2D} \xi(t), \label{eq:zetaode}
\end{equation}
where again, $\tilde{\eta} = 2b/a$. 

Thus, we have fully reduced the dynamics of the system to a single time-varying quantity $\zeta$, which again can be thought of as the \textit{characteristic distance} of the system. Although a considerable number of reductions have been made, the physical behavior of the system is still recoverable by recalling that the instantaneous mean cargo velocity of the system can be recovered from \eqref{eq:hatv}. In other words, $\zeta(t)$ is a proxy for $\hat{v}$, which is the biophysical quantity of interest.

\section{Results}

\subsection{Force Delay Approximation}
\begin{figure}[!h]
\centering{
\includegraphics[width=4in]{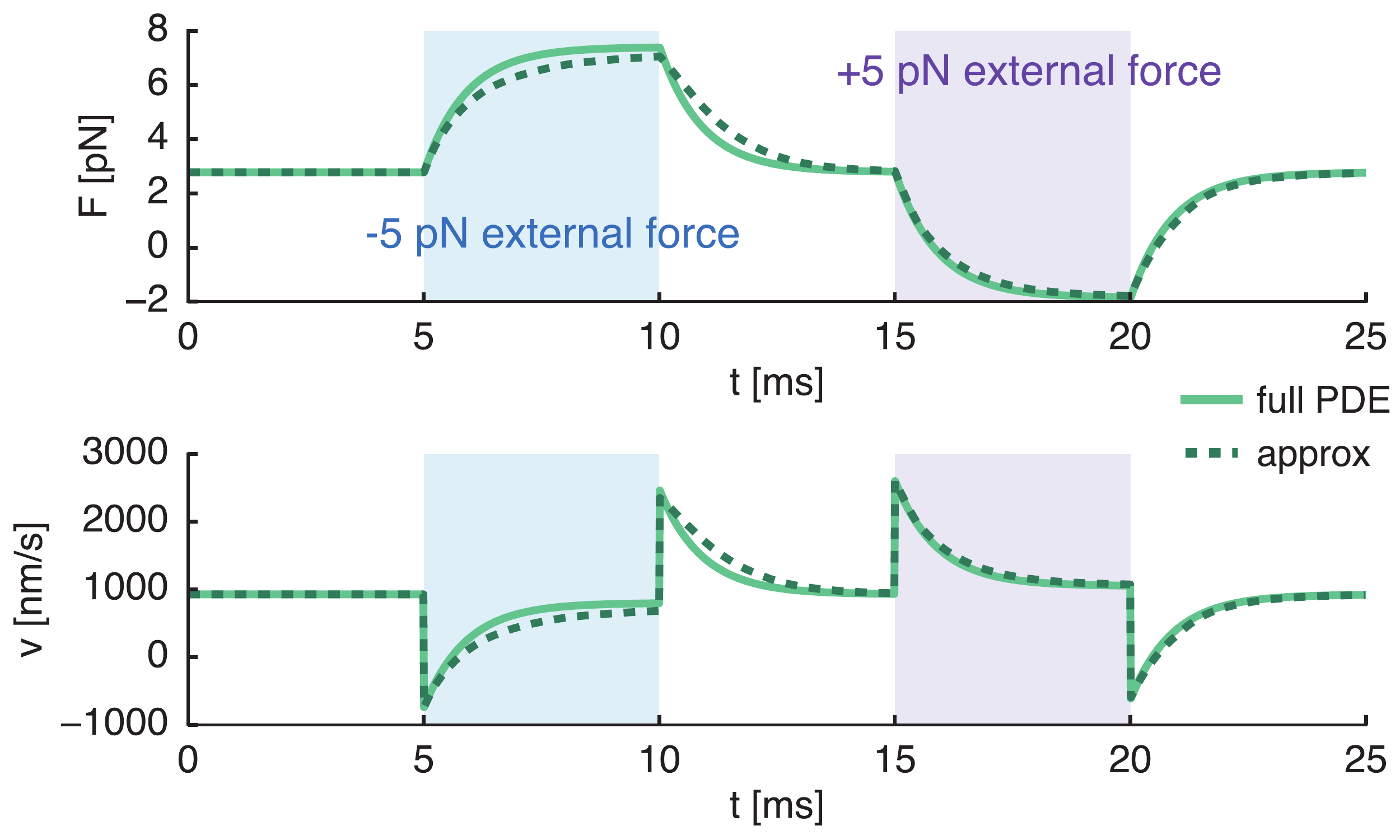}} \\
\caption{{A numerical comparison of the forces and cargo velocity generated by the full mean field motor model (\ref{eq:mean_field}, \ref{eq:instantF}) with  the ``characteristic distance'' approximation described by (\ref{eq:mv}) for one motor population and no thermal noise. In both models, the evolution of the cargo velocity is described by (\ref{eq:v_orig}). External forces are applied to the cargo and removed to illustrate the ability of the reduced model to respond to temporal changes in force.}}\label{fig:approx_compare}
\end{figure}

{To evaluate the validity of the delay approximation inspired by the Ornstein-Uhlenbeck process, we perform numerical simulations of the full mean-field model \eqref{eq:mean_field},\eqref{eq:instantF} and the reduced model \eqref{eq:mv}. For clarity, we simulate only a single motor population (+ direction) and no thermal noise. The approximation fundamentally is one of how motors (and the force generated by them) respond temporally, so a numerical experiment is performed by applying instantaneous external forces to both the mean field model of motors and the reduced model, both of which have cargo dynamics determined by \eqref{eq:v_orig}. Both models are started at the completely unloaded state and run to equilibrium. Once at equilibrium, a -5 pN force (and later +5 pN) external force is applied to the cargo for 5 ms and then removed. The mean-field PDE was simulated using a Lax-Wendroff scheme and the remaining ODEs are computed using a Runge-Kutta 4(5) scheme. The dynamics of the force generated by the motor population and the resulting cargo velocity are tracked and shown in \textbf{Figure \ref{fig:approx_compare}}.}

From \textbf{\ Figure \ref{fig:approx_compare}} we are able to make a number of observations about the validity of the ``characteristic distance'' approximation. For one, the equilibria of the full model and reduced are the same, which is intuitive, given that the reduced model is built from the equilibrium of the full model, as described in \eqref{eq: hatF}. From this, and that the mean-field model can be thought of as a long-time average, we can conclude that there is agreement on long time scales. As the  external force changes instantaneously, both models behave (quantitatively and qualitatively) similarly regardless of the directionality of the force, and therefore, suggests there is also agreement on short time scales. Other external inputs (e.g. sinusoid) were also investigated and yielded similar results. Thus, we have collapsed the force generated by the PDE mean-field description of motors (\ref{eq:mean_field}, \ref{eq:instantF}) into an ODE \eqref{eq:mv} in a ``characteristic distance'' variable and the approximation appears to be valid. 

\subsection{Metastable Behavior}

We perform simulations of \eqref{eq:zetaode} with the parameters specified in \textbf{\ Table \ref{tab:params_general}} with the Euler-Maruyama scheme \cite{Kloeden1992}. The results of a typical simulation can be seen in \textbf{\ Figure \ref{fig:traj_example}}. From this simulation, we see a curious behavior: the characteristic distance $\zeta$ switches between two configurations, or is said to be \textit{metastable}. Elaborating on this, $\zeta$ takes on values near some particular point and then, due to the noise of the system, randomly switches to values centered around another point. The histogram of $\zeta$ values during the simulation, which can also be seen in \textbf{\ Figure \ref{fig:traj_example}} is clearly bimodal, which is a characteristic sign of metastability. Although the two peaks in the figure appear different, this is a consequence of the short time for which the simulation was performed. If more switches were recorded, the two peaks of the histogram would be identical due to the symmetric population assumption, however this time frame was chosen to demonstrate the time-scale on which switching occurs.

\begin{figure}[h]\label{fig:traj_example}
\centering{
\includegraphics[width=5in]{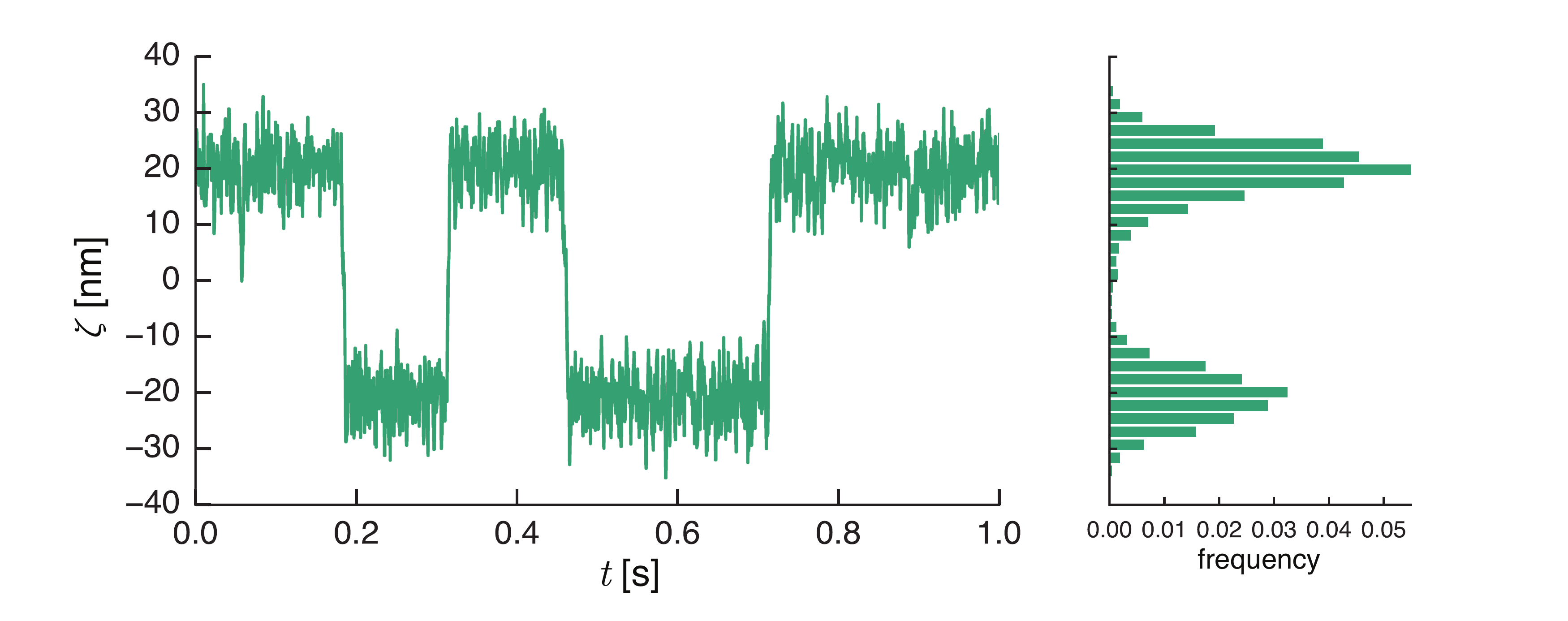}} 
\caption{\protect\textbf{Left:} A typical simulation of \protect\eqref{eq:zetaode} performed with the Euler-Maruyama scheme. The system notably switches between two configurations. \protect\textbf{Right:} A histogram of the values of the simulation, which demonstrates bimodality.\label{fig:traj_example}}\label{fig:traj_example}
\end{figure}

The metastable behavior of the system is apparent from simulations, but can be further elucidated. To do so, consider the corresponding Fokker-Planck equation to \eqref{eq:zetaode}, which describes the probability density $p(\zeta, t \, | \zeta_0, 0)$. That is, the probability density of \eqref{eq:zetaode} given that it started at $\zeta_0$, which is described by  
\begin{equation}
	\partial_t p = -\partial_\zeta \left \{ A(\zeta) p \right \} + 4D \partial_{\zeta \zeta} p, \label{eq:1FPFE}
\end{equation}
where we are abbreviating
\begin{equation}
	A(\zeta) \coloneqq -a\zeta - \frac{2}{\gamma}\left[F_1\left(\frac{\zeta+\tilde{\eta}}{2}\right)+F_2\left(\frac{\zeta-\tilde{\eta}}{2}\right) \right], \qquad D \coloneqq \frac{k_B T}{\gamma}. \label{eq:AU}
\end{equation}

\begin{figure}[h]
\centering{
\includegraphics[width=\textwidth]{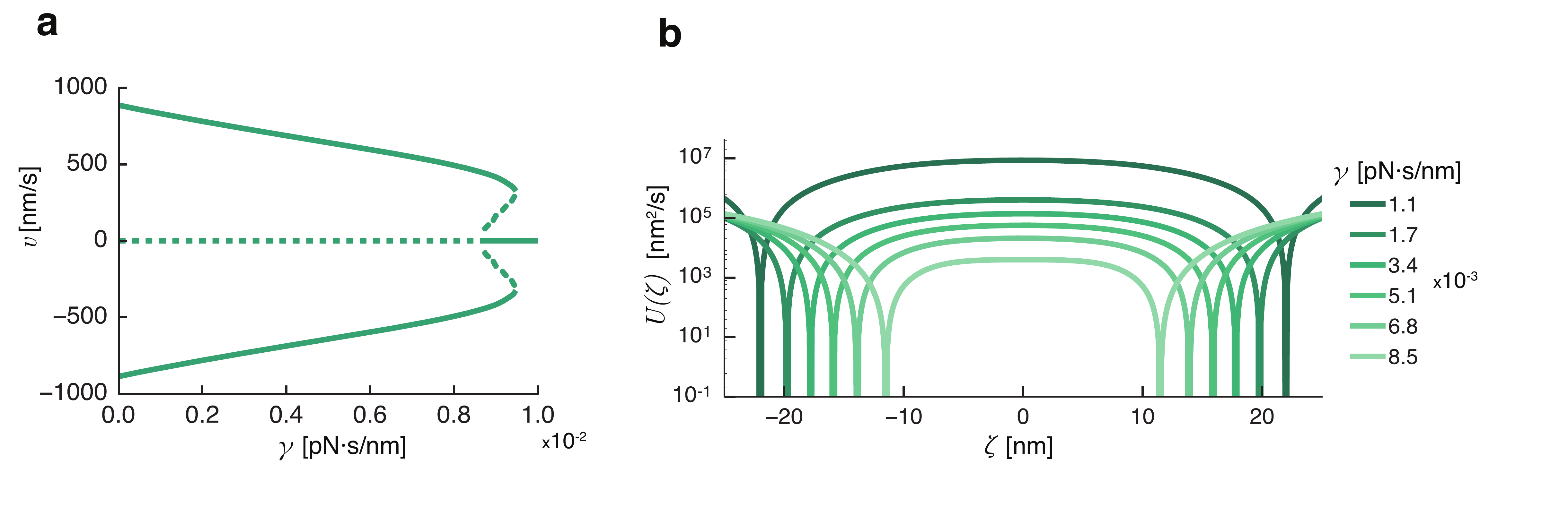}} 
\caption{\protect\textbf{Left:} A bifurcation diagram (as a function of the cargo drag, $\gamma$) for the system, which is computed from the equilibria of \protect\eqref{eq:AU} and then translated into mean cargo velocities by \protect\eqref{eq:hatv}. Dotted lines correspond to unstable equilibria and solid lines are stable. In a wide range of $\gamma$, the system demonstrates a stable positive and negative velocities, or bidirectional motion. \protect\textbf{Right:} the double-well  potential structure \protect\eqref{eq:Upot} as a function of the drag coefficient $\gamma$. As $\gamma$ decreases, the wells get steeper and farther apart.  \label{fig:vp}}
\end{figure}

A bifurcation diagram of the equilibria of $A(\zeta)$ is constructed by varying $\gamma$, the drag coefficient. Since $\zeta$ is not the physical quantity of interest, we translate the equilibria of $\zeta$ into the corresponding mean cargo velocity $\tilde{v}$ under the transformation described by \eqref{eq:hatv}. The resulting bifurcation diagram can be seen in \textbf{Figure \ref{fig:vp}a}. This figure captures exactly the phenomenon described as \textit{bidirectional motion} \cite{Hancock2014}. We see that for a robust range of $\gamma$, the system is \textit{bistable}: there are stable positive and negative mean cargo velocities, which we will denote $v_+, v_-$ respectively. In this same regime, the zero velocity $v_0$ is unstable. Interestingly, in small window of $\gamma$ values, the system is actually \textit{tristable}: two new equilibria emerge in a bifurcation and cause $v_0$ to turn stable. This may correspond to the experimental observation \cite{Kunwar2011} that the system can spend long periods of time in a ``pause'' state, also noting that this same experimental work suggests velocities that agree with those predicted by our model. For large values of $\gamma$, the system only has one stable equilibrium, $v_0$. 

From, \textbf{Figure \ref{fig:vp}a}, the tristable region in $\gamma$-space is fairly narrow. It is possible that other parameters (or more detailed functional forms) would allow for this region to be more robust, but this is not observed. For this reason, we instead focus our study toward the bistable region, where we study the time to switch between the positive and negative velocities. Then, the corresponding potential can be defined by 
\begin{equation}
	U(\zeta) \coloneqq - \int A(\chi) \, \dd \chi. \label{eq:Upot}
\end{equation}
This potential $U(\zeta)$ can be plotted as a function of $\gamma$ in the bistable region of \textbf{Figure \ref{fig:vp}a} and the result is seen in  \textbf{Figure \ref{fig:vp}b}. From the figure, we see that $U(\zeta)$ is a double-well potential. That is, there are two distinct well locations and a peak in the center, all three of which are roots of $A(\zeta)$. Denote the two well locations (stable fixed points of $A(\zeta)$) as $\zeta_{S1}$ and $\zeta_{S2}$, where $\zeta_{S1} < \zeta_{S2}$ and the middle peak (a hyperbolic fixed point of $A(\zeta)$) as $\zeta_{H}$. 

The effect of the drag coefficient $\gamma$ on the potential is non-trivial. Particularly, as $\gamma$ decreases, the wells of the potential $U(\zeta)$ deepen and split farther apart, which alone would suggest an increase in time to switch. However, we later see that there is a counteracting effect in the strength of diffusion.

\subsection{Mean First Passage Time Analysis}

One natural quantity to study in bidirectional systems is the time to switch directions, or the reversal time. Because of the double-potential well structure, this can be thought of as the mean time from one of the metastable points to the hyperbolic point, from which the system relaxes quickly to the other metastable point.  Due to the symmetric motor population assumption, the time to switch states is independent of state. Thus, without loss of generality, we compute the mean first passage time from $\zeta_{S1} \to \zeta_{S2}$ where, again, $\zeta_{S1} < \zeta_{H} < \zeta_{S2}$. 

The analysis of a mean first passage time in a one-dimensional potential is classical \cite{Gardiner2009, Bressloff2014} and is briefly summarized here. Define $G(z,t)$ to be the probability that the system described by \eqref{eq:1FPFE} is in the leftmost potential well at time $t$ given the initial state $p(\zeta,0) = z$. That is, the survival probability density is described by
\begin{equation}
G(z,t) \coloneqq \int_{\zeta_{S1}}^{\zeta_H} p(\zeta, t \, | \, z, 0) \, \dd \zeta. 
\end{equation}
Then, let $T(z)$ define the random variable describing the exit time from this potential well , which satisfies
\begin{equation}
\mathbb{P}\left[T(z) \leq t\right] =  1 - G(z,t) \label{eq:T_PDF}. 
\end{equation}
Taking a derivative of \eqref{eq:T_PDF} yields the density for exit time $f(z,t)$
\begin{equation}
f(z,t) = -\partial_t G(z,t) = -\int_{\zeta_{S1}}^{\zeta_H} \partial_t p(\zeta, t \, | \, z, 0) \, \dd \zeta. 
\end{equation}
From this, we can define the mean first exit time from the potential well, starting at the point $z$ by
\begin{equation}
\tau(z) \coloneqq \langle T(z) \rangle = \int_0^\infty t f(z,t) \, \dd t = \int_0^\infty G(z,t) \, \dd t. \label{eq:tau_def}
\end{equation}
The survival probability $G(z,t)$ satisfies the backward Fokker-Planck equation \cite{Gardiner2009}, which we can integrate and use \eqref{eq:tau_def} to yield the governing equation for the mean exit time density of the system starting at $\zeta_0 = z$, which is
\begin{equation}
	A(z) \tau' + 4D\tau'' = -1, \qquad \tau(\zeta_H) = 0, \qquad \tau'(\zeta_{S1})=0. \label{eq:MFPT_eq}
\end{equation}
The reflecting boundary at $\zeta_{S1}$ is a consequence of starting the system in the well corresponding to this point, as any excursions to the left will quickly relax back to the bottom of the well. The exit location, the hyperbolic point $\zeta_H$, is an absorbing state due to the fast relaxation to the other potential well once the system transverses the peak between them. 

The boundary value problem \eqref{eq:MFPT_eq} does not appear to be solvable analytically due to the complexity of the force curves. However, $\tau(z)$ can be computed numerically in a straightforward manner (in a single integration) by exploiting the linearity of the system. Alternatively, a deep-well approximation can be made for the potential and the classical \textit{Arrhenius formula} can be used to approximate the mean first passage time. For details on both of these methods, see \textit{\ref{S:A_MFPT}}.

The two aforementioned techniques of evaluating $\tau(\zeta_{S1})$ are computed and compared against Monte Carlo simulations of \eqref{eq:zetaode}, again using the Euler-Maruyama scheme, where switching is considered passing the hyperbolic point. The result of these techniques can be seen in \textbf{Figure \ref{fig:mfpts}}. From this, we see that the shooting technique agrees with Monte Carlo simulations and the deep-well approximation is, although qualitatively similar, an overestimate of the switching time. This result is intuitive, as in reality, the wells may not be sufficiently deep for the approximation to work well and therefore allow escape much faster. 

\begin{figure}[h]
\centering{
\includegraphics[width=4.15in]{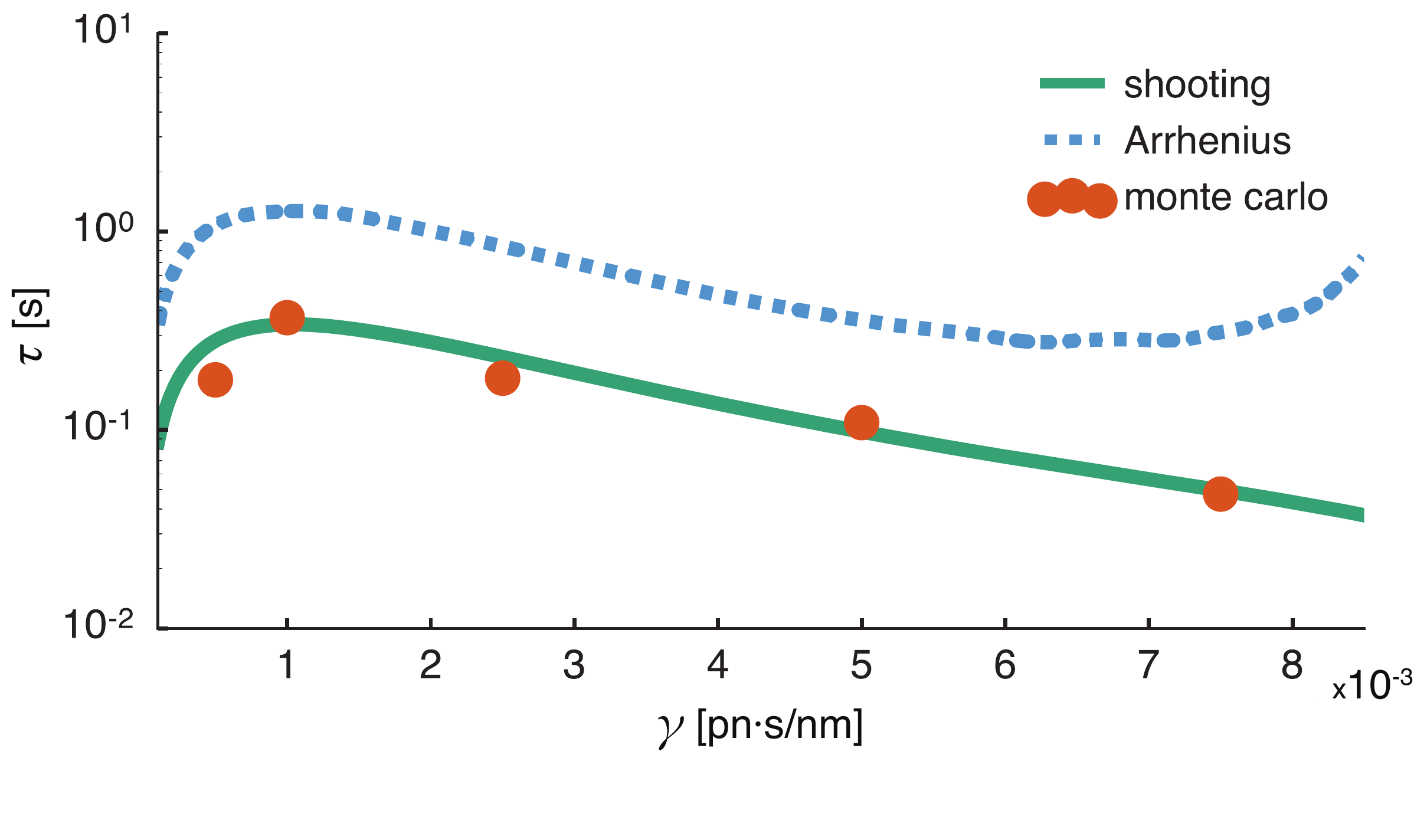}}
\caption{Mean first passage times corresponding to the cargo switching directions. Two approaches to solving \protect\eqref{eq:MFPT_eq} are illustrated: a shooting technique, and the deep-well Arrhenius approximation. The results of an Euler-Maruyama simulation of \protect\eqref{eq:zetaode} are also shown, where switching is considered passing through the hyperbolic point. \label{fig:mfpts}}\label{fig:mfpts}
\end{figure}

The behavior of the mean first passage times as a function of the drag, $\gamma$ is quite interestingly, non-monotonic. That is, as the drag coefficient increases (which can be thought of as the cargo increasing in size), the time to switch initially goes up, but then ultimately goes back down. Mathematically, this complexity stems from $\gamma$ scaling both the potential and the diffusion strength differently, explicitly in \eqref{eq:AU}. As $\gamma$ decreases, the potential wells deepen and spread apart as $\sim1/\gamma$, but the strength of diffusion simultaneously scales by $\sim \sqrt{1/\gamma}$, which are competing effects for the switching time. The resulting behavior is therefore a complex competition between the scaling of the potential and the noise strength, which produces non-monotonicity. Switching due to motor binding and unbinding is \textit{not} expected to demonstrate this same non-monotonicity, as this is a feature of the mismatched scaling in the strength of the driving noise source (diffusion) and the depth of the potential wells. In other words, $\gamma$ does not scale the driving noise source the same way for motor binding dynamics. In other theoretical works that compute the switching time, monotonicity is seen \cite{Guerin2011,Guerin2011a}. 

From a biophysical perspective, it should be noted that the predicted mean first passage times are on the order of $\sim 0.5$[s], which agrees with experimentally observed values \cite{Kunwar2011}. This agreement supports the hypothesis that cargo diffusion is the noise source for bidirectionality. The non-monotonicity of the curve also provides a testable experimental prediction. That is, bidirectional motion via molecular motors could be observed for different cargo drag values (which, could be obtained by varying bead size). If the resulting  mean time to switch directions is found to be non-monotonic, this would further strengthen our theory that cargo diffusion, not motor binding dynamics are indeed the noise source of bidirectionality.

\section{Discussion \& Conclusion}

In this work, we have proposed a mean-field, unequally distributed load description of motor-mediated transport. To understand the behavior of this complex model, we perform a series of reductions. The first, inspired by a simple Ornstein-Uhlenbeck process, quantifies the delay in which motors are able to respond to instantaneous changes in the cargo velocity. Secondly, we use the small mass of the cargo to perform an adiabatic (quasi-steady state) reduction of the system. Due to the correlated noise structure, the final system describes the dynamics of a single ``characteristic distance" which is a proxy for the instantaneous cargo velocity. This resulting stochastic dynamics are observed to be ``metastable'', switching between two distinct states exclusively due of cargo diffusion. These states are associated with positive and negative cargo velocities, meaning the system is bidirectional. To quantify the reversal time of the system, a mean-first passage time analysis is performed and the results are explored as a function of the cargo drag, an experimentally tunable quantity. We find that the predicted switching time agrees with experimental values and also has a non-monotonic dependence on cargo drag, a claim that can be experimentally verified.

The Ornstein-Uhlenbeck analysis for quantifying the ability of a motor to react to instantaneous changes in cargo velocity is of interest in other recent work \cite{Bouzat2016} and in general, causes issue in any work that seeks to use a force-velocity relationship (which is inherently a steady-state analysis) to infer dynamics. In \cite{Bouzat2016}, the authors hypothesize a ``motor memory'' and conclude that models only agree with experimental values appropriately if the motors react to a windowed-time-average velocity. We have quantified this ``memory'' directly from a  physiological description of a single motor on a single run. However, our analysis was only performed for a single motor and was assumed to hold for a population with binding and unbinding. Thus, establishing this reaction for a whole population more explicitly is still desirable and may relate to the results of \cite{mckinley2012b}.

In \cite{Bouzat2016}, the authors also cite the importance of cargo diffusion in models producing results that match experimental values. In our work, we have further illustrated the importance of cargo diffusion by illustrating its ability to produce qualitative changes in motor-mediated transport. Specifically, the fundamental noise driving switching in our model is cargo diffusion, unlike previous unequally distributed load  models which depended on a discrete motor description. This raises the possibility of the importance of diffusion in other aspects of motor-mediated transport.

Thus, we have illustrated that common features of previous work: discreteness of the motors, asymmetry of motor populations, equally distributed loads are \textit{not} necessary to produce a physiologically reasonable model of bidirectional motor transport. This raises uncertainty of which key ingredients may be essential for tug-of-war, making it even more difficult to compare to the alternative regulatory hypothesis of bidirectionality. However, we have provided an experimentally testable prediction of the reversal time as a function of the drag coefficient, which can be tuned by the bead size in experimental setups. If indeed thermal noise is the driver of this switching, then agreement with this experiment would help strengthen the validity of this theory since this feature is not expected from motor binding dynamics as the driving noise source.

\section{Acknowledgments}

This research was partially supported by NSF grant DMS 1122297 and DMS-RTG 1148230.

 \appendix

\section{Ornstein-Uhlenbeck Mean Evolution}
\label{sect:A:OU}

In this section, we show that if the advection term of an Ornstein-Uhlenbeck has a time dependence, a differential equation can be obtained for the mean of the process, demonstrating an effective delay.  

Consider a Fokker-Planck equation of the form
\begin{equation}
\partial_t p = - \partial_x \left [\left\{w(x) - v(t)\right\} p\right] + D \partial_{xx}p. \label{eq:OUreduceFPE}
\end{equation}
Denote $\mu(t)$ to be the mean of the process, that is $\mu = \langle p \rangle$. Then, we have:
\begin{equation}
\dot {\mu} = \deriv{}{t} \int_{-\infty}^{\infty} x \, p(x,t) \, \dd x = \int_{-\infty}^{\infty} x \partial_t p \, \dd x.
\end{equation}
However, we can use \eqref{eq:OUreduceFPE} to find that
\begin{equation}
\dot{\mu} = -\int_{-\infty}^{\infty}  x \partial_x  \left[\left\{w(x) - v(t)\right\} p \right] \, \dd x + \int_{-\infty}^{\infty} x D \partial_{xx}p   \, \dd x,
\end{equation}
which, after integration by parts, yields
\begin{equation}
\dot{\mu} = \langle w(x) \rangle - v.
\end{equation}
Jensen's inequality states that for a convex $w$
\begin{equation}
\langle w(x) \rangle \geq w(\langle x \rangle),
\end{equation}
however, if we assume $w(x)$ is \textit{linear} (as we have done in the model), then Jensen's inequality attains equality and the result is
\begin{equation}
\dot{\mu} = w(\mu) - v(t).
\end{equation}

\section{Methods for 1D MFPT Problems}
\label{S:A_MFPT}

For the sake of generality, consider the one dimensional SDE 
\begin{equation}
	\dd x = A(x) \dd t + \sqrt{2B(x)} \, \dd W,
\end{equation}
which has a corresponding Fokker-Planck equation 
\begin{equation}
	\partial_t p = -\partial_x \left \{A(x) p \right\} + B(x) \partial_{xx}p.
\end{equation}
We are assuming that $A(x)$ has three fixed points, two stable and one hyperbolic, which we'll denote $x_S$ and $x_H$.

 We are then interested in the mean first passage time starting from a point $y$, which we'll denote $\tau(y)$, which satisfies
 \begin{equation}
 	A(y)\tau' + B(y)\tau'' = -1, \qquad \tau'(x_S) =0, \qquad \tau(x_H) =0. \label{eq:BVP}
 \end{equation}

\subsection{Shooting Method}

In this section, we exploit the linearity of \eqref{eq:BVP} to construct a numerical shooting method for constructing a solution. First, we write the system as a first order system, by taking $\sigma = \tau'$, meaning we have 
\begin{equation}
\begin{bmatrix} \tau' \\ \sigma' \end{bmatrix} + \begin{bmatrix}0 & 1 \\ 0 & \frac{A(x)}{B(x)}  \end{bmatrix} \begin{bmatrix}\tau \\ \sigma \end{bmatrix}  = \begin{bmatrix}0 \\ -\frac{1}{B(x)}  \end{bmatrix}, \qquad \begin{bmatrix} \tau(x_H) \\ \sigma(x_S) \end{bmatrix} = \begin{bmatrix} 0 \\ 0 \end{bmatrix} \label{eq:firstorder}.
\end{equation}
To construct a solution to \eqref{eq:firstorder}, we obtain two solutions of initial value problems of the same form and utilize the linearity of the equation to solve the boundary value problem via superposition. Thus, consider the following two systems:
\begin{align}
\begin{bmatrix} p_1' \\ p_2' \end{bmatrix} + \begin{bmatrix}0 & 1 \\ 0 & \frac{A(x)}{B(x)}  \end{bmatrix} \begin{bmatrix}p_1 \\ p_2 \end{bmatrix}  = \begin{bmatrix}0 \\ -\frac{1}{B(x)}  \end{bmatrix}, \qquad \begin{bmatrix} p_1(x_S) \\ p_2(x_S) \end{bmatrix} = \begin{bmatrix} 0 \\ 0 \end{bmatrix} \\
\begin{bmatrix} q_1' \\ q_2' \end{bmatrix} + \begin{bmatrix}0 & 1 \\ 0 & \frac{A(x)}{B(x)}  \end{bmatrix} \begin{bmatrix}q_1 \\ q_2 \end{bmatrix}  = \begin{bmatrix}0 \\ 0  \end{bmatrix}, \qquad \begin{bmatrix} q_1(x_S) \\ q_2(x_S) \end{bmatrix} = \begin{bmatrix} 1 \\ 0 \end{bmatrix}
\end{align}
We now claim $\Upsilon = \begin{bmatrix}\tau  & \sigma \end{bmatrix}^T$ is a linear combination of $P=\begin{bmatrix} p_1 & p_2\end{bmatrix}^T$ and $Q = \begin{bmatrix}q_1 & q_2\end{bmatrix}^T$. In other words, there exists some $\gamma$ such that $\Upsilon = P + \gamma Q$. The value of $\gamma$ is to be determined by making sure the right boundary condition is satisfied
\begin{equation}
\tau(x_H) = p_1(x_H) + \gamma q_1(x_H) = 0 \implies \gamma = -\frac{p_1(x_H)}{q_1(x_H)}.
\end{equation}
Thus, our mean first passage time from $x_S\to x_H$ is then
\begin{equation}
\tau(x_S) = p_1(x_S) + \gamma q_1(x_S) = \gamma. 
\end{equation}
It is worth noting that this actually only requires a \textit{single} ODE integration, as $Q$ is identically constant by construction with $q_1 \equiv 1$ and $q_2 \equiv 0$, and consequently
\begin{equation}
	\gamma = -p_1(x_H).
\end{equation}

\subsection{Arrhenius (Deep Well) Approximation}

The deep-well approximation is a classical technique used to approximate the solution to a mean-first passage time boundary value problem for a double-well potential. Here, we briefly summarize the result but additional details can be found in \cite{Gardiner2009,Bressloff2014}. Define the potential function $U'(y) \coloneqq A(y)$, so $U = \int A(y) \, \dd y$, then, after using an integrating factor and assuming $B$ is constant for simplicity, we have
\begin{equation}
	\tau = \frac{1}{B} \int_{x_S}^{x} e^{U(x')/B}\, \dd x' \int_0^{x'} e^{-U''(x'')/B}\, \dd x''.
\end{equation}
Assuming the potential is deep-welled, the first integral is dominated around the region $x'' = x_S$ and the second dominated by $x' = x_0$, which allows the limits to be changed with small error and therefore can be approximated by
\begin{equation}
	\tau = \frac{1}{B}  \left[\int_{-\infty}^{\infty}e^{-U(x'')/B} \, \dd x'' \right] \left[e^{U(x')/B} \, \dd x'\right].
\end{equation}
Using the method of steepest descent (or simply, Taylor expansion), we finally have the classical \textit{Arrhenius formula}
\begin{equation}
	\tau \sim \frac{2\pi}{\sqrt{|U''(x_H)|U''(x_S)}}e^{(\Delta U)/B}, \qquad \Delta U \coloneqq U(x_H) - U(x_S).
\end{equation}

\printbibliography

\pagebreak 

\title{\textbf{\LARGE \sffamily  Supplemental Information}\\ 
\textbf{\normalsize \sffamily  Bidirectionality From Cargo Thermal Fluctuations in Motor-Mediated Transport}
}
\author{\sffamily Christopher E. Miles, James P. Keener}
\date{\vspace{-.1in}}
\maketitle
\setcounter{equation}{0}
\setcounter{figure}{0}
\setcounter{table}{0}
\setcounter{section}{0}
\makeatletter
\renewcommand{\thesection}{S\arabic{section}}
\renewcommand{\theequation}{S\arabic{equation}}
\renewcommand{\thefigure}{S\arabic{figure}}

\section{Steady-State Force Density}
\label{sect:A:Analytical}

In this section, we construct an analytical solution to the steady-state mean field equation \eqref{eq:mean_field_ss} with the particular choice of functional forms described in the chapter. Thus, we are looking at equations of the form

\begin{equation}
\partial_x \left \{ (w(x) -v ) m \right\} + k_{\text{off}}e^{k |x|/F_D}m = \left \{M - \int_{-\infty}^\infty m(x) \, \dd x  \right\} k_{\text{on}} \delta(x).
\end{equation}
The first observation that can be made is: due to the linearity of this equation, we can reduce it to the study of the simpler equation
\begin{equation}
\partial_x \left \{ (w(x) -v ) u \right\} + k_{\text{off}}e^{k |x|/F_D} u =  k_{\text{on}} \delta(x), \label{eq:uss}
\end{equation}
where $m(x)$, the original solution can be recovered via the relationship
\begin{equation}
m(x) = \frac{M}{1+ U} u(x), \qquad U \coloneqq \int_{-\infty}^{\infty} u(x) \, \dd x.
\end{equation}
We now divide everything through by $k_{\text{off}}$ in \eqref{eq:uss} and recall $w(x) = -ax + b$. Denote the rescaled variables $\cdot/k_{\text{off}}$ by  $\tilde{\cdot}$ and also abbreviate $k/FD = \alpha$, yielding
\begin{equation}
\partial_x \left \{ ( -\tilde{a} x + \tilde{b} - \tilde{v} ) u \right\} + \exp\{\alpha |x|\} u =  \tilde{k} \delta(x). \label{eq:scaledu}
\end{equation}
We can now split this into two scenarios: to the left of $x=0$ and to the right:
\begin{equation}
\begin{cases}
\partial_x \left \{ ( -\tilde{a} x + \tilde{b} - \tilde{v} ) u_L \right\} + \exp\{-\alpha x \} u_L =  0 & \text{ for } x <0,\\
\partial_x \left \{ ( -\tilde{a} x + \tilde{b} - \tilde{v} ) u_R \right\} + \exp\{\alpha x\} u_R =  0 &  \text{ for }x > 0. \label{eq:ucases}
\end{cases}
\end{equation}
These two equations must satisfy a matching condition at $x=0$, so consider integrating \eqref{eq:scaledu} a tiny window around $x=0$ from $-\eps$ to $\eps$, yielding
\begin{equation}
\int_{-\eps}^{\eps} \partial_x \left \{ ( -\tilde{a} x + \tilde{b} - \tilde{v} ) u \right\} + \exp\{\alpha |x|\} u = (b-v) \left [u_R(0) - u_L(0) \right] \quad = \quad  \int_{-\eps}^\eps \tilde{k} \delta(x) \, \dd x = \tilde{k}.
\end{equation}
In other words, we have the matching condition 
\begin{equation}
(b-v) \left [u_R(0) - u_L(0) \right] = \tilde{k}.
\end{equation}
Integrating \eqref{eq:ucases}, we find
\begin{subequations}
\begin{align}
u_L(x) =  \frac{\alpha_L}{\tilde{a}x-\tilde{b}+\tilde{v}} \exp \left \{ \frac{1}{\tilde{a}}  \exp\left(\frac{(-\tilde{b}+\tilde{v})\alpha}{a}\right) \operatorname{Ei}\left(- \frac{(-\tilde{b}+\tilde{v}+\tilde{a}x)}{\tilde{a}}\right) \right\}\\
u_R(x) =  \frac{\alpha_R}{\tilde{a}x-\tilde{b}+\tilde{v}} \exp \left \{ \frac{1}{\tilde{a}} \exp\left(\frac{(\tilde{b}-\tilde{v})\alpha}{a}\right) \operatorname{Ei}\left(\frac{(-\tilde{b}+\tilde{v}+\tilde{a}x)}{\tilde{a}}\right) \right\}, 
\end{align}
\label{eq:uLuR}
\end{subequations}
where $\alpha_R, \alpha_L$ are unknown constants and $\operatorname{Ei}$ is the exponential integral. The matching of these two can be simplified by the realization: only one of $u_L, u_R$ is non-zero. 

That is, if $-\tilde{a}x + \tilde{b} - v >0$, then the advection is rightward (only starting from $x=0$) and therefore $u_L=0$. Similarly, if the advection is leftward then $u_R=0$ necessarily. It should also be noted that \eqref{eq:uLuR} demonstrate the integrable singularity at $x^\star= \frac{-\tilde{v}+\tilde{b}}{\tilde{a}}$, beyond this point, the solution is also necessarily zero. Thus, the solution reduces to either the interval $[0, x^\star ]$ or $[x^\star,0]$ depending on the sign of $x^\star$, or really, if $b>v$.

Thus, if $b>v$, then $x^\star >0$ and $u_L <0$ and if $b < v$ then $x^\star <0$ and $u_R =0$. Thus, if $b>v$, then our matching condition provides us $\alpha_R$:
\begin{equation}
\alpha_R = -\tilde{k} \exp \left \{ \frac{1}{\tilde{a}} \exp\left( \frac{(\tilde{b}-\tilde{v})\alpha}{\tilde{a}}\right) \operatorname{Ei}\left( \frac{(-\tilde{b}+\tilde{v})\alpha}{\tilde{a}} \right)\right\}.
\end{equation}
Similarly, in the case that $b<v$, we have 
\begin{equation}
\alpha_L = \tilde{k} \exp \left \{ \frac{1}{\tilde{a}} \exp\left( \frac{(-\tilde{b}+\tilde{v})\alpha}{\tilde{a}}\right) \operatorname{Ei}\left( \frac{(\tilde{b}-\tilde{v})\alpha}{\tilde{a}} \right)\right\}.
\end{equation}
Thus, we have constructed all components of the analytical solution to the original steady state equation.
 
 \clearpage 

\section{Force-Velocity Curves}

In this section, we plot the steady state force-velocity curves described by \eqref{eq:vtilde}. In these plots, the parameter values are taken to be those described by \textbf{\sffamily Table \ref{tab:params_general}} except for one parameter (shown in the legend), which is adjusted over a range of values.
\begin{figure}[H]
  \centering
        \includegraphics[width=2.25in]{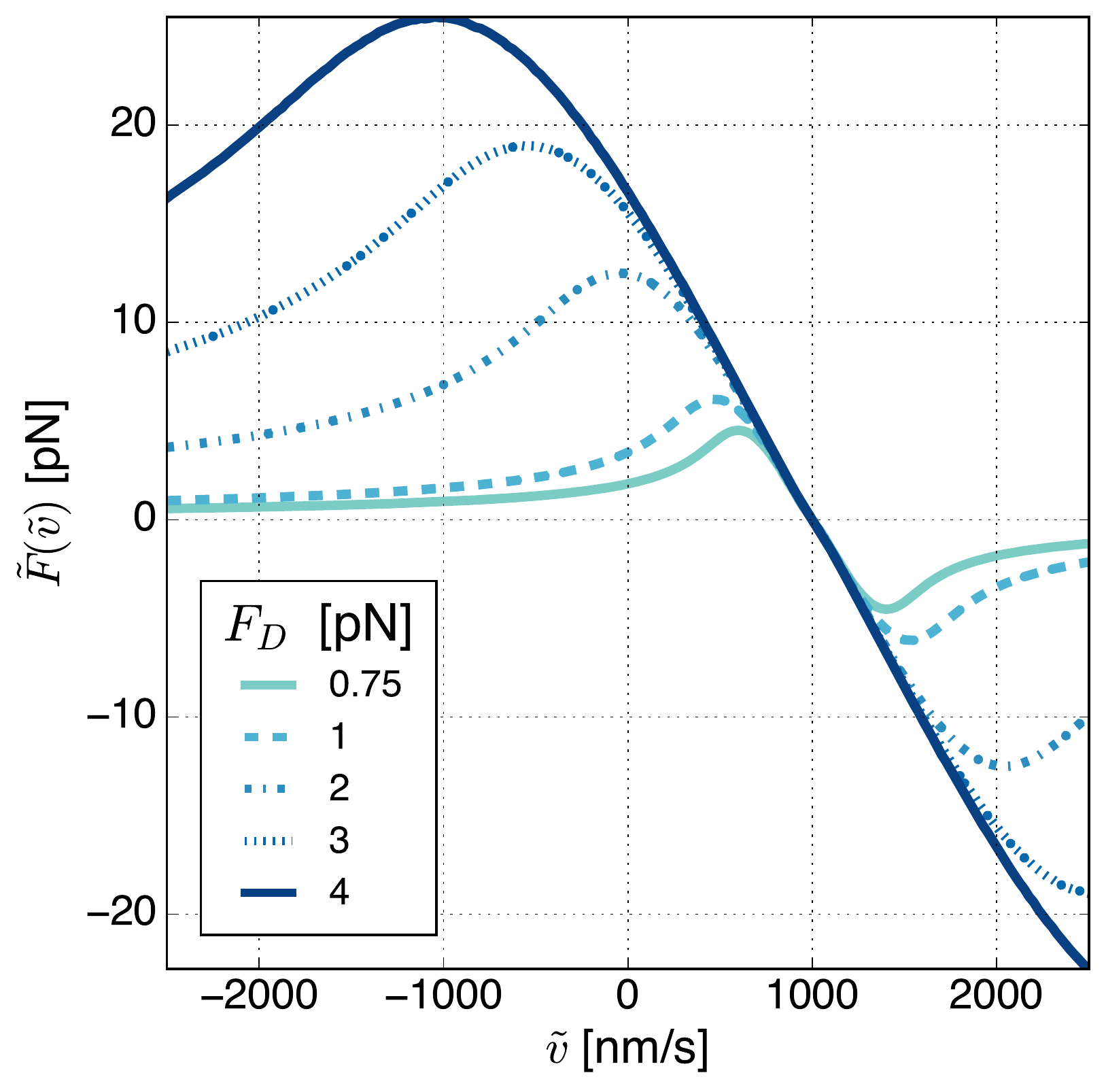}
    ~ 
        \includegraphics[width=2.25in]{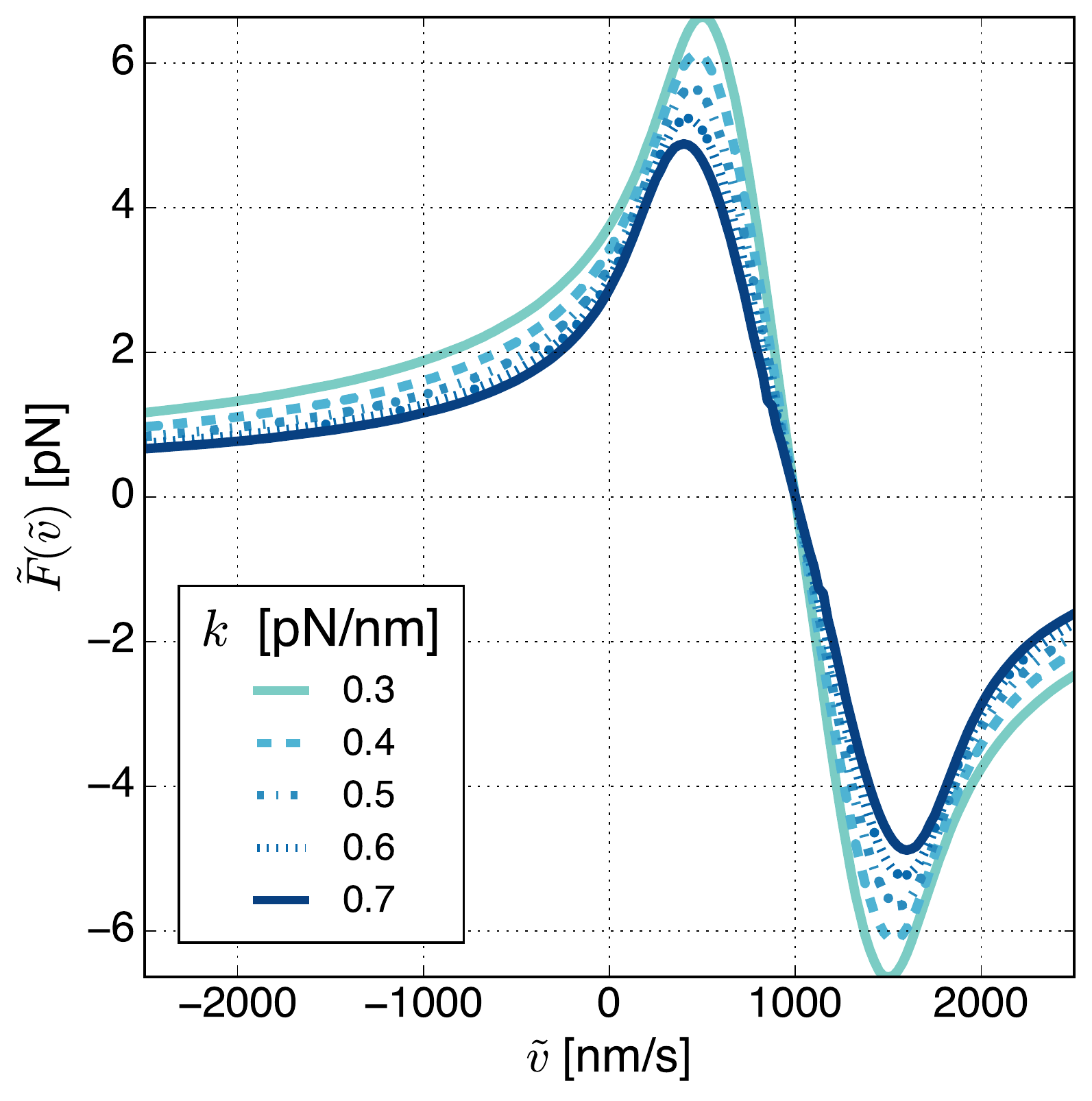}

        \includegraphics[width=2.25in]{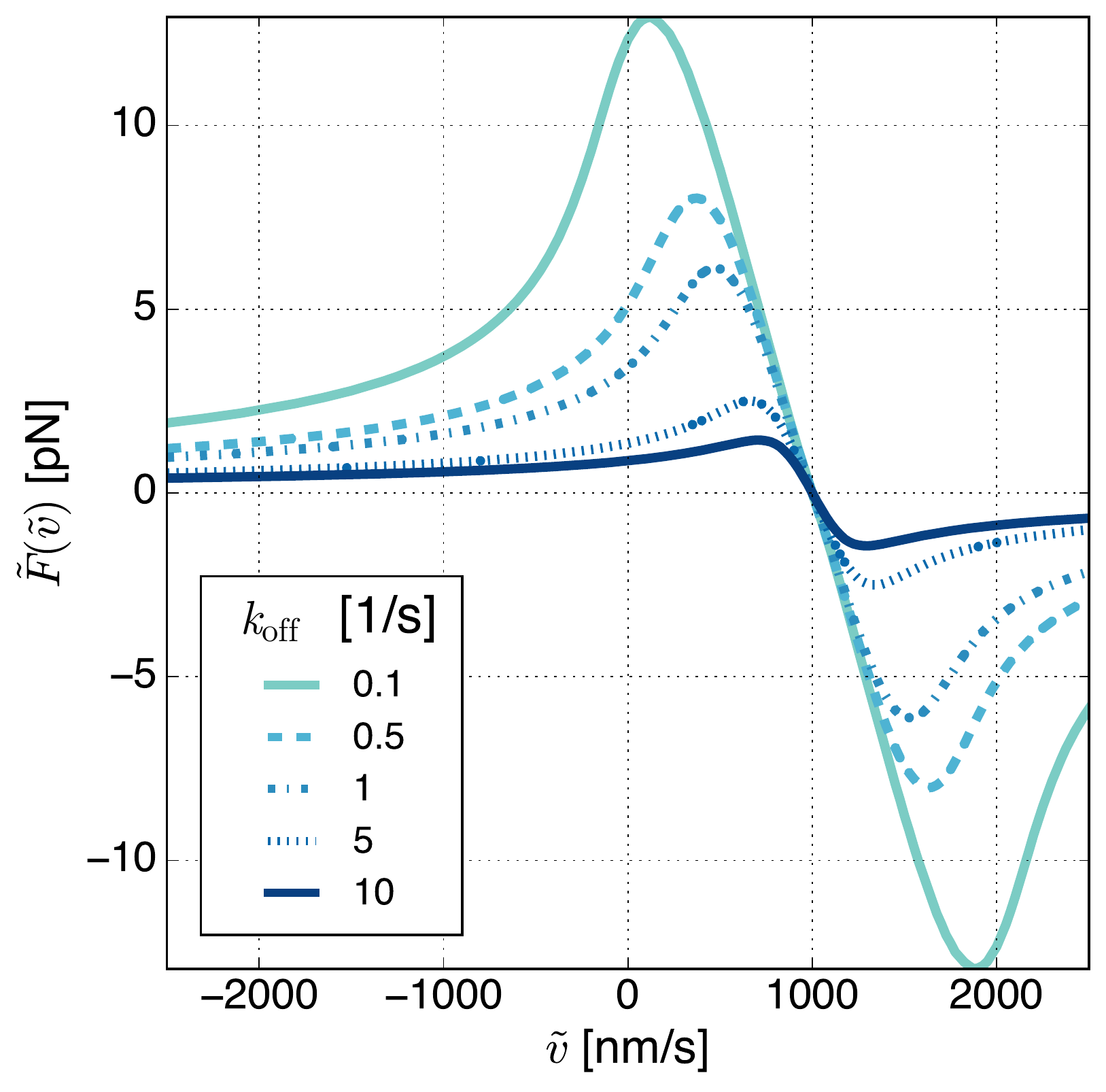}
        ~
        \includegraphics[width=2.25in]{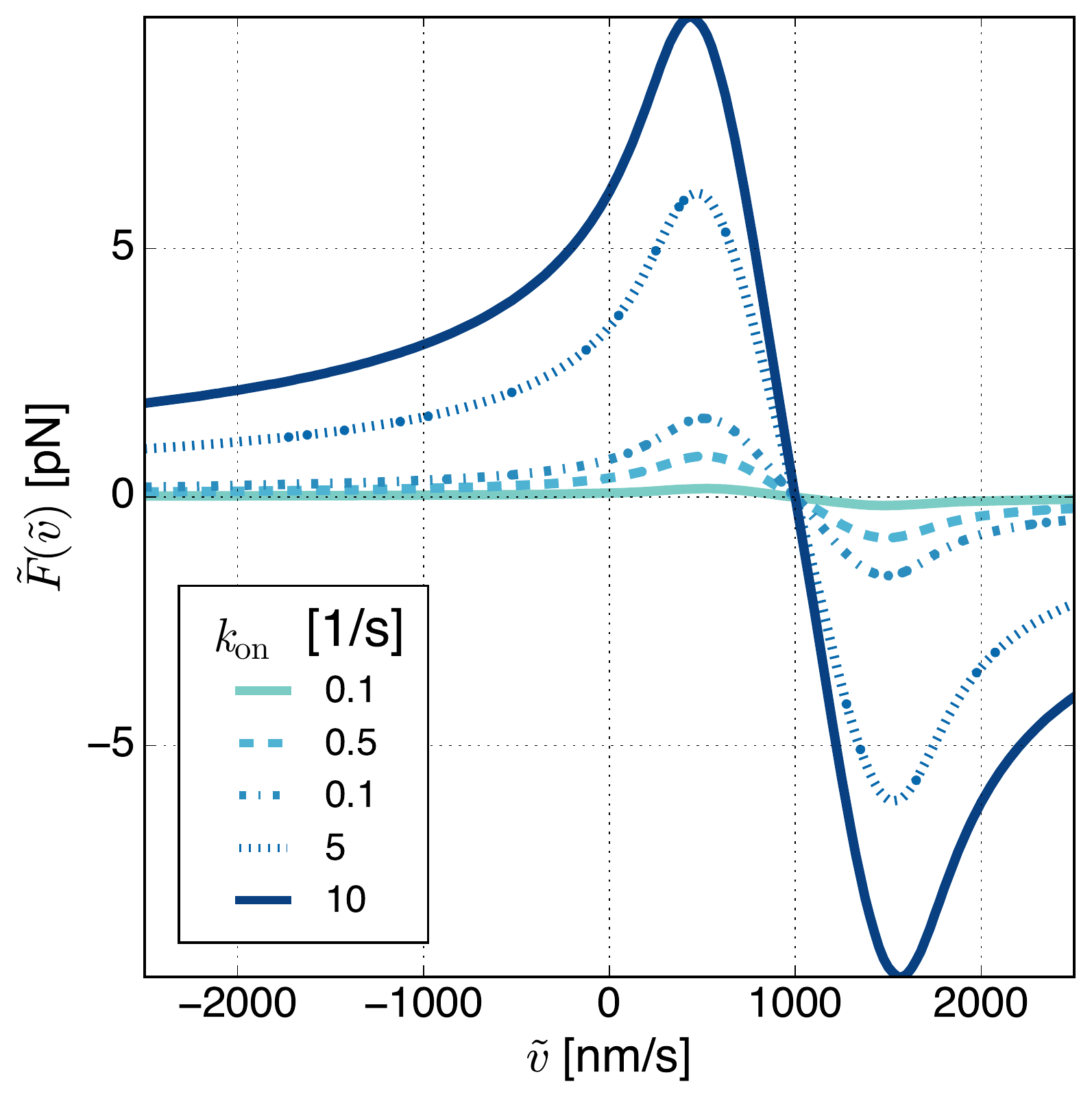}

        \includegraphics[width=2.25in]{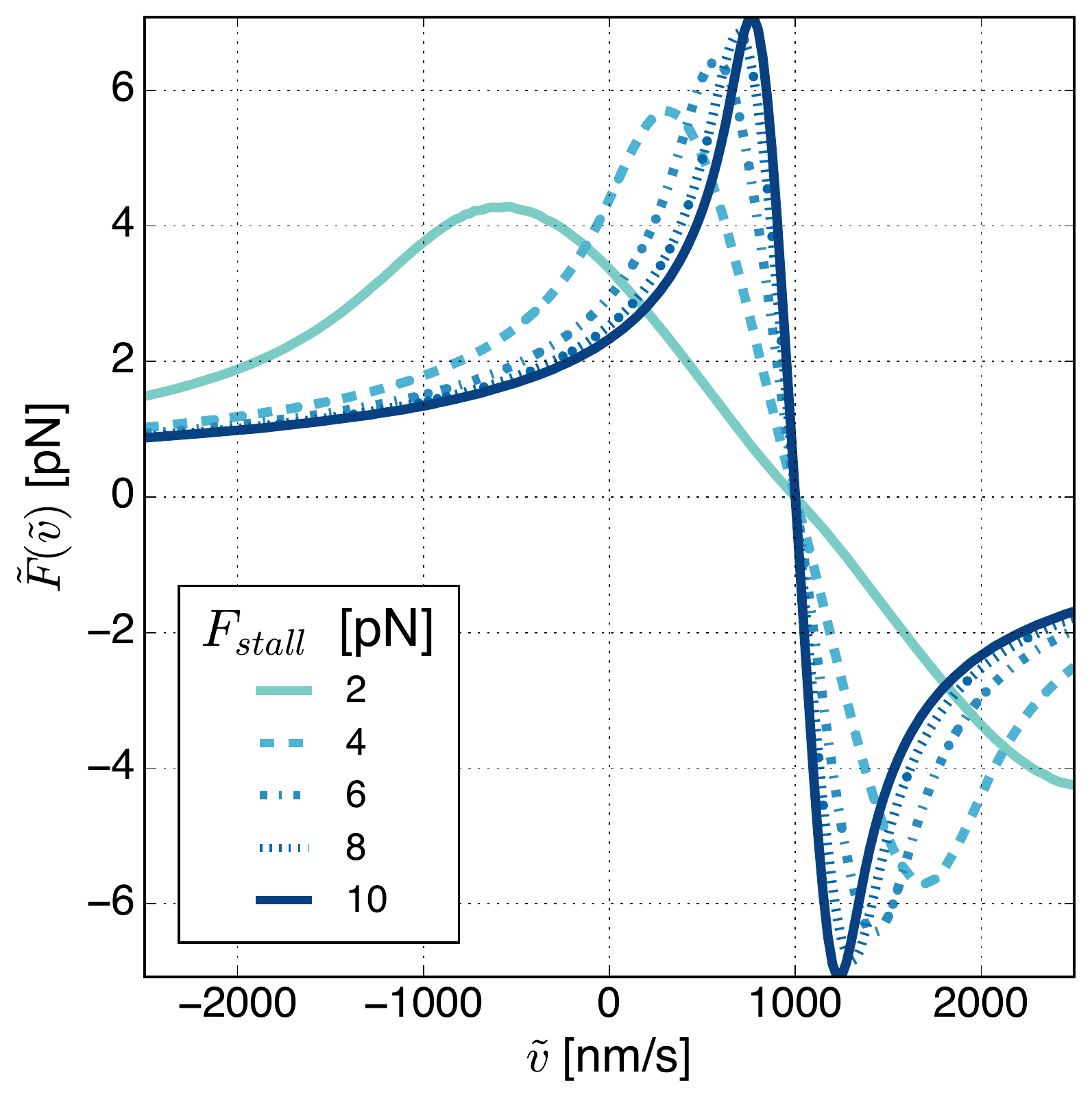}
    ~ 
        \includegraphics[width=2.25in]{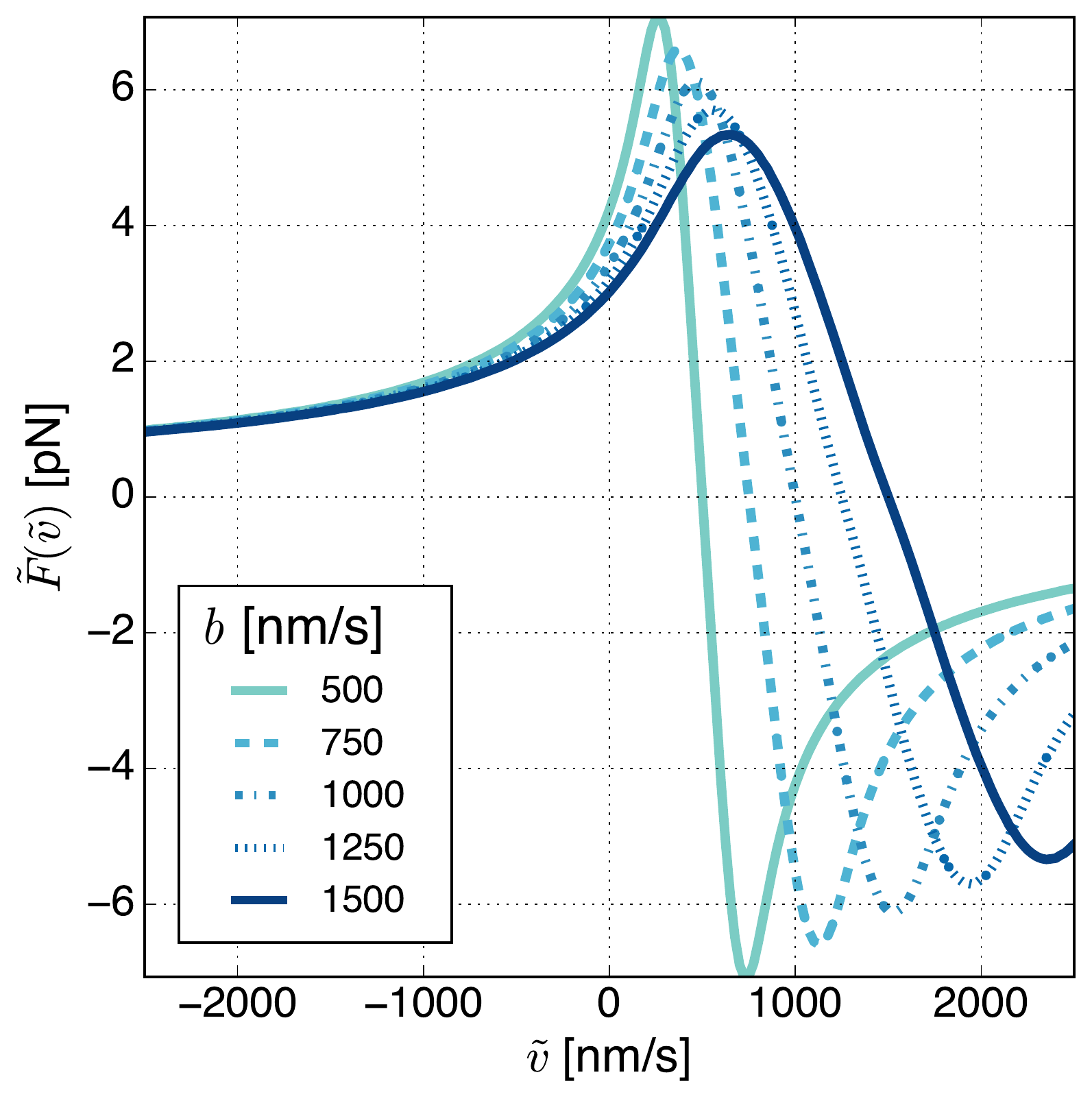}

\caption{Plots of the steady state force distribution $\tilde{F}$ parameterized by the velocity of the cargo for different parameter values.}
    \label{fig:Ftildeplots}
    \end{figure} \clearpage
\section{Adiabatic Reduction Details}
\label{sect:S:adiabatic}
In this section, we perform an adiabatic reduction of \eqref{eq:mv}, which, recalling the form of $w(x)$ and using $F(x) = kx$ for the sake of illustration, yields
\begin{equation}
\mathcal{M} \dot{v} + \gamma v = k x + \sqrt{2 \gamma k_B T} \xi (t), \qquad \dot{x} = ax + b - v,
\end{equation}
which is equivalent to the Fokker-Planck equation
\begin{equation}
\pderiv{p}{t} = - \pderiv{}{x} \left \{ (ax + b - v) p\right\} -  \frac{1}{\mass} \pderiv{}{v} \left \{( k x - \gamma v)p \right\} + \frac{k_B T \gamma }{\mass^2} \pderiv{{}^2p}{v^2}. \label{eq:adiaFPE}
\end{equation}
We first perform a non-dimensionalization. Let $y = x/x_0, \tau = t/t_0$, which provides a scaling on the velocity $u = v t_0 / x_0$, all of which are dimensionless, where we particularly take $t_0 = \gamma/k$, and set $\gamma t_0/\mass =1/ \varepsilon$, which gives us that $\gamma^2/k\mass = 1/\varepsilon$. We can then also set the last term $\gamma k_B T t_0^2/ \mass^2 x_0^2 = 1/\varepsilon$, which gives us that $x_0 = \sqrt{k_BT\gamma^2/\mass k^2}$. Then, \eqref{eq:adiaFPE} becomes 
\begin{equation}
\pderiv{p}{\tau}  = - \pderiv{}{y} \left \{ \left (\alpha y + \beta - u \right) p \right \} + \frac{1}{\varepsilon}\pderiv{}{u} \left \{ (u-y) p + \pderiv{p}{u} \right\},  \label{eq:adiaFPEreduce}
\end{equation}
which we denote
\begin{equation}
\pderiv{p}{\tau} = \frac{1}{\varepsilon} \mathbb{L}_1p + \mathbb{L}_2 p.
\end{equation}
Note, the null-space of the fast operator, $\mathbb{L}_1$ is \textbf{not} the same as the classical Brownian due to a different choice of $\eps$.

Now, if $\phi \in \operatorname{null} (\mathbb{L}_1)$, then it satisfies the following differential equation:
\begin{equation}
\pderiv{\phi}{u} + (u-y)\phi = 0,
\end{equation}
which has a solution
\begin{equation}
\phi(u) = \frac{1}{\sqrt{2\pi}} \exp \{-(u-y)^2/2\}.  \label{eq:gaussian}
\end{equation}
Define the projection operator $\mathbb{P}$ as
\begin{equation}
\mathbb{P}f \coloneqq \phi(u,y) \int_{-\infty}^{\infty} f(u,y) \, \dd u, \qquad \Q \coloneqq 1-\mathbb{P}.
\end{equation}
We then split our solution $p$ into the part in the null-space of the fast operator and otherwise. That is,
\begin{equation}
p = \mathbb{P}p + \Q p = v + w,
\end{equation}
where we take $v$ to be of the form $v = f(y,t) \phi(u,y)$, as it is in the null space of $\mathbb{L}_1$, and $f$ is some unknown amplitude. 

We first consider applying $\mathbb{L}_2$ to $v$ for later calculations
\begin{align*}
\mathbb{L}_2 v = \mathbb{L}_2 \mathbb{P}p & = - \pderiv{}{y} \left \{ \left (\alpha y + \beta - u \right) f(y) \phi(u,y) \right \}.
\end{align*}
Now, applying $\mathbb{P}$ to this result yields
\begin{equation}
\mathbb{P}\mathbb{L}_2 \mathbb{P} p = -  \pderiv{}{y} \left\{ (\alpha y  + \beta - y) f \right \} \phi(u,y). 
\end{equation}
Next, we consider applying $\mathbb{P}$ and $\Q$ to the Fokker-Planck equation to yield the differential equation
\begin{equation}
\mathbb{P}\left ( \pderiv{p}\tau \right) = \pderiv{v}\tau = \mathbb{P}\left(\frac{1}{\varepsilon}\mathbb{L}_1 + \mathbb{L}_2\right)(v+w) = \mathbb{P} \mathbb{L}_2 v+ \mathbb{P}\mathbb{L}_2 w = - \pderiv{}{y} \left\{ (\alpha y  + \beta - y) f \right \} \phi + \mathbb{P}\mathbb{L}_2 w,
\end{equation}
based on the first calculation and the fact that $\mathbb{P}\mathbb{L}_1 =0$ by construction. Next, we have
\begin{align}
\Q \left(\pderiv{p}\tau\right) = \pderiv{w}\tau &= \Q\left(\frac{1}{\varepsilon}\mathbb{L}_1 + \mathbb{L}_2\right)(v+w)\\
&= \frac{1}{\varepsilon}\mathbb{L}_1 w + \Q \mathbb{L}_2(v+w) \\
&=  \frac{1}{\varepsilon}\mathbb{L}_1 w +  \mathbb{L}_2 v + \mathbb{L}_2 w - \mathbb{P}\Ltwo v - \mathbb{P}\mathbb{L}_2 w. 
\end{align}
Noting that, again $\mathbb{P}\mathbb{L}_1 = 0$ and $\mathbb{L}_1 v =0$ by construction. We now take $w$ to be in quasi-steady state, meaning it must satisfy
\begin{equation}
  \frac{1}{\varepsilon}\mathbb{L}_1 w = -\mathbb{L}_2 v + \mathbb{P}\Ltwo v,
\end{equation}
which, when using the definitions of these operators yields
\begin{align}
\frac{1}{\varepsilon} \pderiv{}{u}\left \{ (u-y)w + \pderiv{w}{u} \right\} =  \pderiv{}{y} \left \{ \left (\alpha y + \beta - u \right) f(y) \phi(u,y) \right \}  -  \pderiv{}{y} \left\{ (\alpha  y  + \beta - y ) f \right \} \phi(u,y).
\end{align}
We integrate once with respect to $u$ to get rid of a derivative on the left hand side, finding that
\begin{equation}
    \frac{1}{\varepsilon} \left \{ (u-y)w + \pderiv{w}{u} \right\} =  \phi \left \{ f(y)(u - \alpha y - \beta ) + f' \right\}.
\end{equation}
and therefore, using an integrating factor
\begin{equation}
w = \frac{\eps}{2} \phi u \left \{ f(y)(u - 2\alpha y - 2\beta) + 2f'(y) \right\}.
\end{equation}
Now, using this form of $w$, we must compute $\mathbb{P}\mathbb{L}_2 w$, since that is the term in the $\partial v/\partial t$ equation. First, applying $\mathbb{L}_2$, by definition:
\begin{align}
\mathbb{L}_2 w &= - \pderiv{}{y} \left \{ \left (\alpha y + \beta - u \right) w(u,y) \right \}.
\end{align}
and now projecting yields
\begin{equation}
\mathbb{P}\mathbb{L}_2 w = \varepsilon \left [f(y) + y f'(y) + f''(y) \right] \phi(u).
\end{equation}
Thus, our differential equation for $v$ is 
\begin{equation}
\pderiv{v}\tau = - \varepsilon \pderiv{}{y} \left\{ (\alpha y  + \beta) f \right \} + \mathbb{P}\mathbb{L}_2 w = - \varepsilon \pderiv{}{y} \left\{ (\alpha y  + \beta) f \right \} + \varepsilon \left [f(y) + y f'(y) + f''(y) \right] \phi(u),
\end{equation}
from which, we can conclude
\begin{equation}
\pderiv{f}{\tau} = - \varepsilon \pderiv{}{y} \left\{ (\alpha y  + \beta) f \right \} + \varepsilon \pderiv{}{y}\left\{ yf(y) \right \} + \varepsilon \pderiv{{}^2f}{y^2},
\end{equation}
in the original variables,
\begin{equation}
\pderiv{f}{t}  = -    \pderiv{}{x} \left\{ \left(ax + b - \frac{k}{\gamma} x \right ) f(x) \right \}  + \frac{k_B T}{\gamma} \pderiv{{}^2 f}{x^2}. \label{eq:adiabatic_result}
\end{equation}

\end{document}